\documentclass[a4paper,fleqn,usenatbib]{mnras}

\usepackage[T1]{fontenc}
\usepackage{ae,aecompl}

\usepackage{graphicx}	
\usepackage{amsmath}	
\usepackage{rotating}
\usepackage{txfonts}

\title[Young and embedded clusters in Cygnus-X]
{Young and embedded clusters in Cygnus-X: \\ evidence for building up the IMF?}

\author[F. F. S. Maia et al.]{F. F. S. Maia$^{1,2,3}$\thanks{E-mail:
francisco.maia@obs.ujf-grenoble.fr}, E. Moraux$^{1,2}$ and I. Joncour$^{1,2}$ 
\\
$^{1}$Univ. Grenoble Alpes, IPAG, F-38000 Grenoble, France \\
$^{2}$CNRS, IPAG, F-38000 Grenoble, France \\
$^{3}$CAPES Foundation, Ministry of Education of Brazil, Bras\'ilia - DF 70040-020, Brazil
}

\date{Accepted for publication in MNRAS}

\pubyear{2015}

\begin{document}
\label{firstpage}
\pagerange{\pageref{firstpage}--\pageref{lastpage}}
\maketitle

\begin{abstract}

We provide a new view on the Cygnus-X north complex by accessing for the first time the low mass content 
of young stellar populations in the region. CFHT/WIRCam camera was used to perform a deep 
near-IR survey of this complex, sampling stellar masses down to $\sim$0.1 M$_\odot$. Several analysis 
tools, including a extinction treatment developed in this work, were employed to identify and uniformly 
characterise a dozen unstudied young star clusters in the area. 
Investigation of their mass distributions in low-mass domain revealed a relatively uniform log-normal IMF 
with a characteristic mass of 0.32$\pm$0.08 M$_\odot$ and mass dispersion of 0.40$\pm$0.06.
In the high mass regime, their derived slopes showed that while the youngest clusters 
(age $<$ 4 Myr) presented slightly shallower values with respect to the Salpeter's, our older clusters 
(4 Myr $<$ age $<$ 18 Myr) showed IMF compliant values and a slightly denser stellar population. 
Although possibly evidencing a deviation from an 'universal' IMF, these results also supports a scenario
where these gas dominated young clusters gradually 'build up' their IMF by accreting low-mass stars formed in 
their vicinity during their first $\sim$3 Myr, before the gas expulsion phase, emerging at the age of $\sim$4 Myr 
with a fully fledged IMF. Finally, the derived distances to these clusters confirmed the existence of at least 3 
different star forming regions throughout Cygnus-X north complex, at distances of 500-900 pc, 1.4-1.7 kpc 
and 3.0 kpc, and revealed evidence of a possible interaction between some of these stellar populations and the 
Cygnus-OB2 association.

\end{abstract}

\begin{keywords}
open clusters and associations: Cygnus-X -- infrared: stars -- stars: pre-main-sequence -- 
 stars: low mass -- stars: luminosity function, mass function
\end{keywords}



\section{Introduction}

The Cygnus X area is one of the richest star forming regions (SFR) known in the Galaxy. It presents an 
extended structure to which are associated numerous optical, infrared and radio objects. Early studies 
divided this region into 9 OB associations \citep{h79} housing over a thousand OB stars and tens of compact 
HII regions and stellar nurseries. Inside these large associations several concentrations have been recognised 
as young stellar clusters or embedded agglomerates \citep{ct01, db01}. 

However, observational efforts to study this region have been hampered by heavy absorption with complex 
patterns caused by a foreground molecular cloud structure known as the Great Cygnus Rift, that also 
appears to harbour star formation \citep{k00}. Moreover, when looking towards Cygnus the local spiral arm 
is seen tangentially, so that structures at different distances superpose in the line of sight, causing 
confusion as to the real spatial structure and content of these associations.

For example, although Cygnus-X is actually divided into two large molecular cloud complexes (Cygnus-X 
north and Cygnus-X south) separated by a large clearing hosting the Cygnus OB2 association, it is still not 
clear wether the north and south cloud complexes are related to each other or to the Cygnus OB2 
association projected between them \citep{kry14}. Recent investigation on the OB stars 
\citep{cpft08,wdm15} indicates that star formation has been ongoing for the last $\sim$10 Myr in the region, 
taking place early in its southern region and then later in Cygnus OB2 and the northern complex, thus 
generating a mixture of different aged populations in the field. However it is not understood if these events 
happened in a triggered manner since the distance to these regions are not well constrained.

In this matter, radio CO imaging of the gaseous content by \citet{s06} indicated that all the structures in 
the Cygnus-X region were interconnected, probably at the same distance of Cygnus OB2. Such findings 
were somewhat supported by maser parallaxes done by \citet[][hereafter R12]{r12} on 5 discrete sources 
across the northern complex, finding a mean distance of 1.4 kpc. However, higher resolution CO 
mapping by \citet[][hereafter G12]{g12} uncovered a much more complex gas structure towards Cygnus-X
north, detecting three distinct layers of CO emission related to (i) the Great Cygnus Rift at 500-800 pc, (ii) 
W75 and DR17 at 1.0-1.8 kpc and (iii) DR21 at 1.5-2.5 kpc.

In this context, the many young star clusters in this region had been largely neglected. These objects 
are excellent probes from which a range of physical parameters can be determined.
Recent work on the characterisation of Cygnus stellar clusters has largely used infrared data from 
the Two Micron All Sky Survey (2MASS) \citep[][hereafter LK02]{bdb03, lk02}, mostly due to its prompt 
availability and ability to probe considerably deeper than optical surveys. It is clear however that
deeper photometry is needed to properly investigate this region. The moderate distance 
(m-M$\approx$11) and high reddening (Ak $\approx$ 3.0) towards this region can make the 2MASS 
sample highly incomplete for stars fainter than A0 spectral type ($\sim$5 M$_\odot$). 

The most comprehensive study of the young clusters in this region has been performed by LK02, who 
detected 11 previously unknown stellar clusters in the area and characterised the morphology and the 
stellar content of total sample of 22 targets. However, due to the limitations of the data, the authors did not 
determined ages nor distances for many of their targets adopting instead the mean values for the 
Cygnus population, thus precluding a global analysis of the region based on the cluster parameters.  

To address this issue, we have acquired deep near-infrared (NIR) images of a large extent of the Cygnus-X
north complex ($\sim$1$^\circ$ $\times$ 1$^\circ$) including a dozen of catalogued clusters and HII regions 
and reaching about 4 magnitudes deeper than previous studies, often sampling stellar masses down to
0.1 M$_\odot$. We have developed a method to constrain most probable age and distance intervals for a 
given stellar population using the statistical distributions of stellar extinction, derived via (stellar)
evolutionary models. Age, distance, extinction, structural parameters and mass distribution were determined
for each catalogued object and used to get an overall view of the Cygnus-X north.

This paper is organised as follows. In Sect.2 we describe the observations and processing of our data 
as well as the photometry and calibration. Sect. 3 describes our cluster sample and Sect. 4 the methodology 
adopted for investigating it. Our results and conclusions are drawn in Sect. 5 and 6 respectively.

\section{Data Handling}

\subsection{Acquisition and processing}

\begin{figure}
\centering
\includegraphics[width=\linewidth]{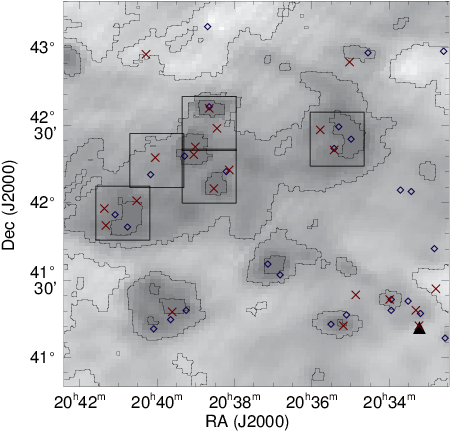}
\caption{IRIS 100$\umu$m dust emission contour map of the region of interest, located $\sim$ 
1.5 deg. northeast of the Cygnus OB2 association centre (black triangle). The observed WIRCam fields 
(squares), known clusters (crosses) and HII regions (diamonds) in the region are also shown. 
 Contour levels are at 150, 300, 750 and 2000 MJy/sr.}
\label{fig:iras}
\end{figure}

The Wide-Field Infrared Camera (WIRCam) at the Canadian-France-Hawaii Telescope (CFHT) was 
employed to observe five fields inside a Cygnus-X cloud complex located about 1.5 deg. to the northeast 
of the centre of the Cygnus OB2 association. An IRIS \citep{ml05} 100$\umu$m dust emission contour 
map of this region is shown in Fig. \ref{fig:iras} depicting the position of the observed WIRCam fields. 
Catalogued stellar clusters and HII regions in the field (see Sect.~\ref{sec3}) are also shown.

Observations were carried out under photometric conditions in six nights between 04/09/12 and 29/10/12, 
yielding seeing values between 0.6\arcsec and 0.9\arcsec. The total integration time in each filter was 
960s in $J$, 1200s in $H$ and 480s in $K_s$, designed as to provide good quality photometry ($S/N 
\approx 25$) down to the following magnitude limits: $J \approx 20.0$, $H \approx 19.5$ and $K_s 
\approx 18.5$. These limits were chosen to sample stars in the sub-solar mass regime, considering the
mean extinction and distance of this cloud complex (LK02). As usual for near-infrared imaging, these 
integration times were fragmented into several short exposures composing a dither pattern with more 
than 20 positions in each field.

Data detrending was carried out at CFHT using the WIRCam pipeline, `I`iwi. Post-processing of the data 
included recalculation of the astrometric solution against the Two Micron All Sky Survey (2MASS) catalogue 
\citep{2mass} with the SCAMP software \citep{bertin06}, and resampling and co-addition of $\sim$500
individual frames into three master mosaics (one in each filter) with the SWARP software \citep{bertin02}. 
The astrometric recalibration was very precise, with RMS residuals about $\sim$0.05\arcsec internally and 
$\sim$0.12\arcsec in relation to 2MASS, well below the WIRCam pixel size (0.3\arcsec). Furthermore, a 
comparative inspection of the images have shown no considerable ($< 10 \%$) degradation of the 
point-spread function (PSF) in the resulting mosaics when compared to the individual frames. Finally, the 
derived map of the PSF FWHM in the resulting mosaics (see next section) showed an smooth solution with 
values ranging from 0.6\arcsec to 1.0\arcsec, consistent with the reported seeing values on the individual 
frames.

\subsection{Photometry}
\label{sect:photometry}

Source extraction and PSF modelling was carried out in the master mosaics using SExtractor and PSFEx 
softwares \citep{bertin11} interactively. Initially, source detection was done at the 2-$\sigma$ level in the 
deeper mosaic ($K_s$) without employing any image filtering, as to maximize the recovery of fainter objects, 
particularly near bright objects. This initial catalogue was filtered out of artefacts, spurious detections and 
extended sources by using morphological parameters derived from the sources isophotal profiles 
(e.g. ellipticity, half-light radius) and local PSF parameters (e.g. FWHM, reduced $\chi^2$) to generate a 
filtered catalogue containing about 3$\times$10$^5$ stars. Finally, Kron-like 
elliptical apertures and PSF fitting were carried out independently in each mosaic ($J$, $H$ and $K_s$) to 
derive fluxes and magnitudes for all sources in the filtered catalogue. 

\begin{figure}
\centering
\includegraphics[width=\linewidth]{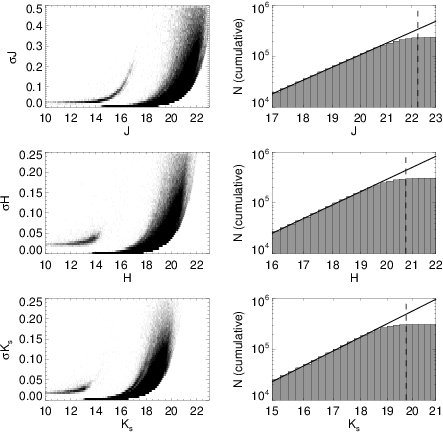}
\caption{Derived photometric uncertainties (left panels) and cumulative source counts (right panels) as 
function of the derived PSF magnitude in filters $J$ (upper panels), $H$ (middle panels) and $K_s$ 
(bottom panels). The brighter uncertainty regime in the left panels are due to recovered 2MASS sources 
(see text). In the right panels, the 50$\%$ completeness level (vertical dashed lines) were estimated from
the extrapolated stellar numbers (solid lines).}
\label{fig:phot}
\end{figure}

The left panels of Fig.~\ref{fig:phot} shows the magnitude uncertainties derived from the PSF photometry in 
each filter. From these values we found that the magnitude limits reached at a $S/N \sim 25$ 
($\sigma$=0.04) were $J \sim 20.5$, $H \sim 20.0$, $K_s \sim 19.0$. The right panels of Fig.~\ref{fig:phot} 
show the cumulative source counts as a function of magnitude. In each diagram, a linear relation was fit to 
the the brighter bins of the distribution allowing us to extrapolate the expected source counts for the fainter 
bins and estimate the source incompleteness as a function of the magnitude. Therefore, we have derived 
the $50\%$ completeness level at: $J_{50}$=22.25, $H_{50}$=20.75, $K_{s\,50}$=19.75 and the $95\%$ 
completeness level at: $J_{95}$=20.75, $H_{95}$=19.50, $K_{s\,95}$=18.50.

\begin{figure}
\centering
\includegraphics[width=\linewidth]{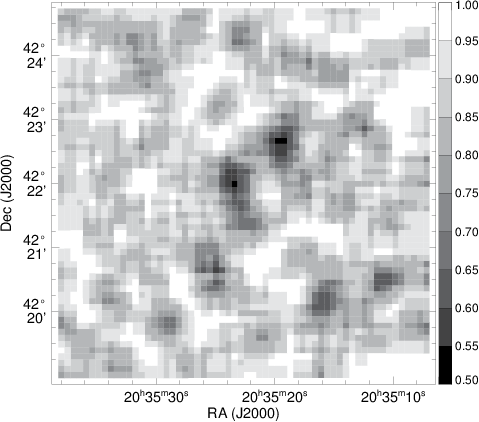}
\caption{Completeness map derived for the cluster LK12, using artificial stars with $K_s$=18.5. 
Although the average completeness value throughout this map was found to be $\sim$86\%, the densest parts 
of the cluster present local completeness values (represented by the colourbar) as low as 50\%. Resolution of 
the map is $\sim$30\arcsec.}
\label{fig:compl}
\end{figure}

While these derived completeness limits should be appropriate for most of the overall field population, higher 
than average incompleteness is expected near the clusters centre, due to crowding effects.
Therefore, a more rigorous completeness estimate was conducted in a large region around each cluster by 
adding artificial stars in a regular grid with a $\sim$6\arcsec spacing, corresponding roughly to the size of the
derived PSF. Local completeness was then estimated by redoing the photometry and calculating the artificial 
stars recovery fraction inside a $\sim$30\arcsec kernel across each cluster region. This process
was repeated several times in each filter, covering a wide range of magnitudes with 0.5 magnitude steps. 
Fig.~\ref{fig:compl} shows the derived completeness map around the cluster LK12 using artificial stars with 
$K_{s\,95}$=18.50, corresponding to the 95\% completeness level determined from stellar counts. It can be 
seen that, even if completeness level averages to $\sim$86\% throughout this cluster region, it can reach 
values as low as 50\% in the more crowded regions near its centre.

\begin{figure}
\includegraphics[width=0.48\linewidth]{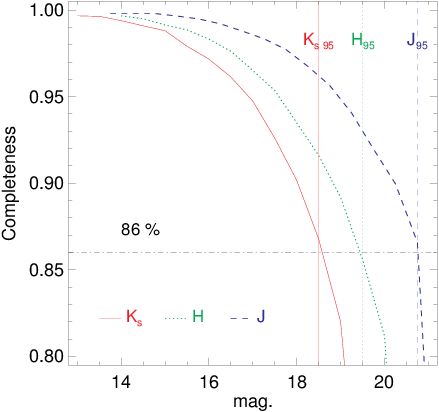} \hskip 0.25cm
\includegraphics[width=0.45\linewidth]{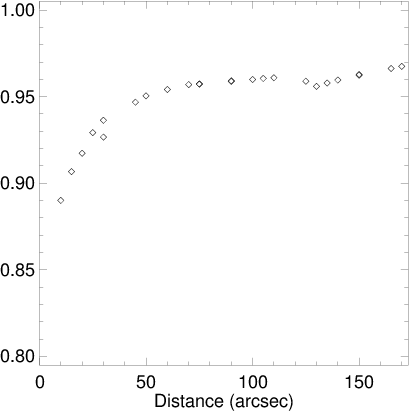}
\caption{{\it Left}: Average completeness around LK12 as function of magnitude, derived from artificial 
star simulations in the $J$, $H$ and $K_s$ bands (dashed, dotted and solid lines, respectively). The 95\% 
completeness derived from stellar counts in each filter (vertical lines) corresponds approximately to a 86\% 
level (horizontal line), as derived from artificial stars simulations. {\it Right}: Integrated completeness across all 
magnitudes fainter than $K_{s\,95}$=18.50, as a function of the distance from LK12 centre. Although 
completeness is lower at the centre of the cluster, it raises in towards the outer regions to approximately the 
95\% value found by stellar counts.}
\label{fig:complcor}
\end{figure}

The average completeness values calculated in those maps as function of magnitude is shown on the left 
panel of Fig.~\ref{fig:complcor}, for cluster LK12. It can be seen that completeness decreases from nearly 
100\% at bright magnitudes to $\sim$86\% at $K_{s\,95}$=18.50, $H_{95}$=19.50 and $J_{95}$=20.75. 
This implies that, although a small correction is still necessary to equalise the
completeness levels as derived from stellar counts and artificial star simulations, they appear to be 
uniformly correlated across all bands. The right panel shows the total completeness as function of 
the distance to cluster centre, obtained by integrating the completeness maps in the $K_s$-band across the 
entire magnitude range tested, down to the $K_{s\,95}$ limit. As it can be seen, it is remarkable that, 
although completeness is lower towards the centre of LK12 it raises quickly to about 95\% in the outer 
regions of the cluster, effectively matching the overall field completeness value found by star counts.

\subsection{Calibration}
Although WIRCam broad-band filters $J$, $H$ and $K_s$ have response curves similar to the 2MASS 
filters, enough differences exists to justify a colour calibration between the two filter sets. Compared
to the equivalent 2MASS filters, the WIRCam ones present a central wavelength that is slightly redder in
the $J$ filter and slightly bluer in the $H$ and $K_s$, while also presenting significantly larger bandpasses 
in the $H$ and $K_s$ filters.

\begin{figure}
\centering
\includegraphics[width=\linewidth]{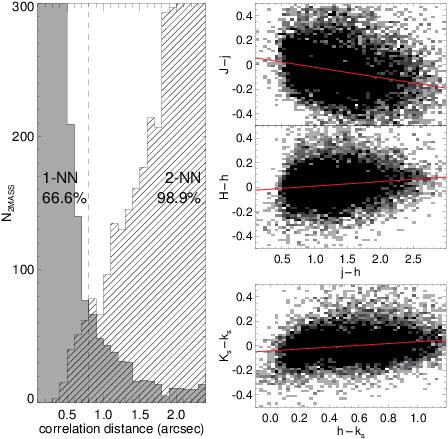}
\caption{Left: distribution of the first (shaded histogram) and second (dashed histogram) nearest neighbour 
of 2MASS sources in our catalogue. The fraction of sources in each distribution with correlation distances 
smaller/larger than the adopted correlation threshold (dashed line) is also shown. Right: photometric 
calibration fits (solid lines) to the 2MASS catalogue in the $J$ (top), $H$ (middle) and $K_s$ filters (bottom).}
\label{fig:calib}
\end{figure}

WIRCam observations usually presents a $\sim$3 magnitude overlap between its 
saturation limit and 2MASS completeness level thus allowing for a generous sample of common stars  
to derive a photometric calibration between the two datasets. In our dataset near 12000 common sources 
were found using a correlation radius threshold of 0.8\arcsec between the datasets, derived from a nearest 
neighbour analysis of 2MASS sources relative to our catalogued sources, shown in Fig. \ref{fig:calib}. 
The adopted correlation distance threshold delimits the regime where most 2MASS sources have 
just one counterpart in our catalogue, shown by the higher number of first neighbour correlations relative to 
second neighbours ones. At greater correlation distances, two catalogue counterparts becomes more likely 
to be found for each 2MASS source, as the number of second neighbour correlations becomes higher. 
Therefore, the determined threshold maximizes the matching of 2MASS sources to our catalogue 
($\sim$$67\%$), while maintaining low levels ($\sim$1$\%$) of possible mismatch due to second 
neighbours.

Selected common sources were used to calculate a simple photometric calibration between the two 
datasets accounting for a zero-point magnitude difference and a linear colour correction according to the 
following relations:
\begin{align}
J-j = &\ j_0  + j_c\,(j-h) \label{eq:calibj} \\ 
H-h = &\ h_0  + h_c\,(j-h)  \label{eq:calibh} \\ 
K_s-k_s = &\ k_0  + k_c\,(h-k_s) \label{eq:calibk}
\end{align}

\noindent
where the lowercase letters represent 2MASS magnitudes, uppercase letters represent magnitudes from 
our catalogue, $j_0$, $h_0$ and $k_0$ are the zero-point magnitude terms and $j_c$, $h_c$ and $k_c$ 
are the colour terms. To minimize the effects of outliers, we have employed a "robust" linear fitting 
method and applied a simple sigma-clipping algorithm at 3-sigma. Fig.~\ref{fig:calib} shows the calibration 
fits of these equations over our derived aperture photometry in each filter. 

Table~\ref{tab:calib} compares our fitted colour terms from aperture and PSF derived photometry with the 
values found by CADC\footnote{http://www.cadc-ccda.hia-iha.nrc-cnrc.gc.ca/en/wirwolf/docs/filt.html}. 
Since uncertainties were not provided for these reference values, we inferred the 
1-sigma deviation based on the FWHM of their provided colour terms distributions$^1$. Considering these 
uncertainties, our aperture derived coefficients showed a good correlation with the reference values for all 
filters, even though the apertures used in each case were different. Although the PSF derived colour terms 
follow the same trends, they are significantly higher than their aperture counterpart. This is most likely due 
to the fact that the profile-fitting photometry is able to recover more flux from the bluer stars used in the 
calibration than the aperture photometry. These stars are usually close to our saturation limits presenting 
an extended light halo that can be easily unaccounted by the small aperture used. The derived zero-point 
magnitude terms were small, with uncertainties inferior to $0.005$ magnitudes in all filters.

\begin{table}
\caption{Photometric calibration colour terms}
\begin{tabular}{l r@{$\pm$}l r@{$\pm$}l r@{$\pm$}l} \hline
  & \multicolumn{2}{c}{aperture} & \multicolumn{2}{c}{PSF} & \multicolumn{2}{c}{CADC$^*$} \\ \hline
$j_c$     & -0.080 & 0.002 & -0.056 & 0.004 & -0.064&0.017  \\
$h_c$    &  0.037 & 0.002 &  0.048 & 0.003 &  0.031&0.009 \\
$k_c$    &  0.075 & 0.003  & 0.093 & 0.004 &  0.058&0.026 \\ \hline
\end{tabular}

$^*$ uncertainties inferred from the colour term distributions
\label{tab:calib}
\end{table}

Since we were mainly interested in recovering bright, saturated stars to our catalogue we have opted to 
calibrate the 2MASS magnitudes into the WIRCam system, instead of the other way around. Apart from 
being simpler as it does not involve inverting the system of equations (\ref{eq:calibj}-\ref{eq:calibk}), it also 
prevents the addition of the calibration errors into our photometry. Therefore, equations 
(\ref{eq:calibj}-\ref{eq:calibk}) were employed directly to 
convert the 2MASS magnitudes of stars brighter than our saturation limit into the WIRCam photometric 
system. Bright stars for which the PSF fitting has not converged were also recovered from the 2MASS
catalogue. Uncertainties due to 
the calibration were properly propagated into these stars magnitude errors; they appear as a brighter, 
narrower strip in the left panels of Fig.~\ref{fig:phot}.

Our final catalogue thus comprises about 310000 stars spanning a 12 magnitudes range in $J$, $H$ 
and $K_s$ filters and reaching $\sim$5 magnitudes deeper than 2MASS. It is available for download at 
Vizier\footnote{http://vizier.u-strasbg.fr/viz-bin/VizieR}.

\section{Cluster identification}
\label{sec3}

The first step in analysing stellar populations is a proper identification of 
cluster candidates as stellar overdensities relative to the local field population, at scale lengths of a few 
parsecs. Confirming the physical nature of the candidate cluster members usually requires further analysis 
using statistical methods, colour-magnitude diagrams (CMD's) and/or spectroscopy.

Fig.~\ref{fig:iras} shows that many of the known clusters in the region (see Table~\ref{tab:catclu})
are often correlated with known HII regions \citep{p03} and/or heavy 100$\umu$m dust emission, indicating 
that these stellar populations might be in their early formation stages. However, since the catalogued clusters 
in this region have never been studied in detail, it is not clear if they even represent coeval physical systems. 
Since our observations are significantly deeper than 2MASS, from which most of these clusters were 
identified, the previously undetected, fainter population can provide a strong constraint in determining the 
physical nature of these objects. 

\begin{figure*}
\centering
\includegraphics[angle=90,width=\linewidth]{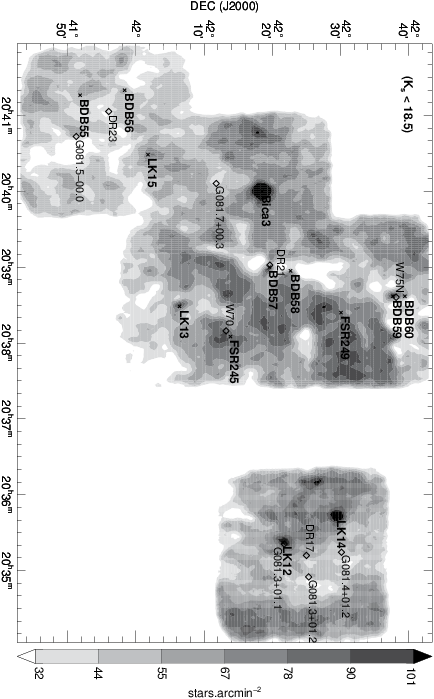} \\ \vspace{3mm}
\includegraphics[angle=90,width=\linewidth]{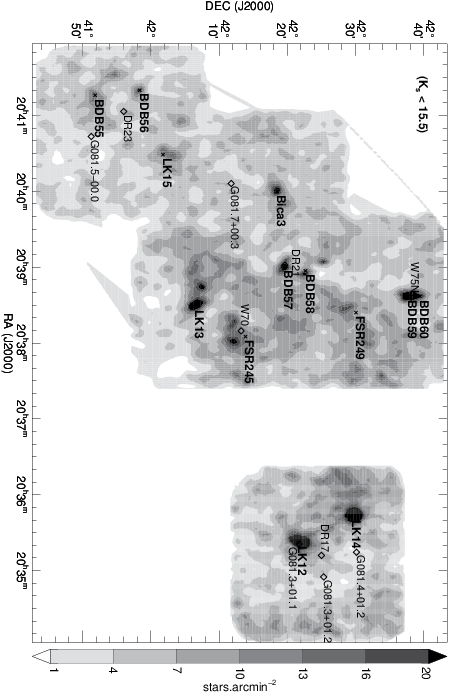}
\caption{Stellar density maps of our final catalogue sampling only stars brighter than $K_s$=18.5 (upper 
panel) and $K_s$=15.5 (lower panel). Density levels are indicated by the colour bar on the right.
Catalogued clusters in the region are labeled in boldfaced text and indicated by crosses. HII regions are
labeled by normal text and indicated by diamonds.}
\label{fig:dmap}
\end{figure*}

To sample the local stellar density through our field of view we have constructed stellar density charts by 
counting detected objects inside a flat circular kernel around each star. This produced an irregular 
grid on the position coordinates which was interpolated into an uniform grid. Since this approach enhances
density features with dimensions similar to the kernel size, we have iterated this procedure with a range of 
different kernel sizes (from 10\arcsec to 90\arcsec). The resulting density charts were then merged by median 
stacking the individual charts into a final density map. Fig.~\ref{fig:dmap} compares two such charts built by 
including all stars brighter than our 95$\%$ completeness level ($K_s < 18.5$) and by including only a 
brighter subset ($K_s < 15.5$). It can be seen that although all catalogued 
clusters correspond to a density nearby enhancement at the brighter subset, some of them are 
overwhelmed by field population when the fainter population is considered, raising doubt as to the real 
physical nature of these objects. This is specially concerning in the identification of new cluster candidates
since other uncatalogued density enhancements that seems distinguishable from the fainter population 
might be in fact artefacts generated by field fluctuations. Table~\ref{tab:catclu} presents the coordinates of the 
known clusters in the region, shown in Figs.~\ref{fig:iras} and \ref{fig:dmap}, with their respective literature 
reference. 

\begin{table}
\caption{Catalogued clusters in the region}
\begin{tabular}{lccr} \hline
 ID    & RA (J2000) & Dec (J2000) & Ref. \\ \hline
LK12   & 20:35:22 & +42:21:36 & 1 \\ 
LK14   & 20:35:43 & +42:29:24 & 1 \\ 
FSR245 & 20:38:05 & +42:13:52 & 2 \\ 
FSR249 & 20:38:24 & +42:30:03 & 2 \\ 
LK13   & 20:38:29 & +42:06:25 & 1 \\ 
BDB60  & 20:38:37 & +42:39:25 & 3 \\ 
BDB59  & 20:38:37 & +42:37:40 & 3 \\ 
BDB58  & 20:38:57 & +42:22:40 & 3 \\ 
BDB57  & 20:39:00 & +42:19:35 & 3 \\ 
Bica3  & 20:40:00 & +42:18:30 & 4 \\ 
LK15   & 20:40:29 & +42:01:48 & 1 \\ 
BDB56  & 20:41:20 & +41:58:22 & 3 \\ 
BDB55  & 20:41:16 & +41:51:51 & 3 \\ \hline
\end{tabular}

$^1$ \citet{lk02} \\
$^2$ \citet{fsr07} \\
$^3$ \citet{bdb03} \\
$^4$ \citet{bbd03} 
\label{tab:catclu}
\end{table}

When completeness is not an issue a real stellar cluster should present an excess of stars relative to the 
field, even when its fainter population is considered. Since we are dealing with a star forming region, we do 
not expect severe dynamical evolution effects that could remove the fainter population of these 
clusters. Heavier than average extinction, however, could efficiently mask the fainter stars in some regions,
making the identification of these objects very susceptible to the larger field fluctuations of the 
fainter magnitude population. Since our completeness levels were derived globally, we can only confirm the 
nature of the emerged, more populous stellar clusters that can be easily identifiable in both magnitude 
limited density charts shown in Fig.~\ref{fig:dmap} and presenting stellar sequences in the CMD that can be 
clearly  differentiated from the field ones. Namely, those are clusters: LK12, LK14, FSR249, LK13, BDB58, 
Bica3, LK15. The other objects will be considered as candidate clusters, pending a more detailed 
investigation on their proposed members.

\section{Physical parameters}
\label{method}

Although a stellar density enhancement over the field is often a prime condition for the identification of a 
star cluster candidate, such enhancement could easily come from statistical fluctuations of the field 
population, inhomogeneous extinction over the line-of-sight or incompleteness effects. Therefore, 
subsequent analysis of the candidate is mandatory to access the physical nature of the stellar group. 

In this sense, we have applied several methods to assess the spatial structure, to derive the local visual 
extinction and to account for the local field population in each cluster candidate, 
in an attempt to uncover their physical parameters. These methods are described below as they are 
applied to LK12, as an example.

\subsection{Structural parameters}

It is known that star clusters undergo a particular structural evolution during their lifetimes, evolving f
rom a very hierarchical to a more globular structure and then expanding as they lose members, towards 
dissolution \citep[e.g][]{pz10}. The timescales involved in such evolution, specially at younger ages, are very 
dependent on the number of stars, their spatial dimension and the gas content, but can be very short 
both locally and globally through the cluster \citep[e.g.][]{k14,gou12}.

To characterise the structural properties or our sample, we have employed fittings of the King-Profile 
analytical function \citep{k62} over magnitude limited radial density profiles (RDP) derived from stellar 
counts. For this purpose, the central coordinates of each object were updated by calculating the 
barycentre of the stellar sample inside 1\arcmin of the peak density value in our star density maps. 
A RDP was then constructed by deriving the mean stellar density inside consecutive 
annular bins with a constant radial bin around the centre coordinates. This procedure was iterated at 
different radial bin values for improved sampling. Lastly, we have used the completeness maps derived 
at Sect.~\ref{sect:photometry} to perform a completeness correction to the derived RDP of our clusters, 
down to the $K_{s\,95}$ limit. (see Fig.~\ref{fig:complcor}).

\begin{figure}
\centering
\includegraphics[width=\linewidth]{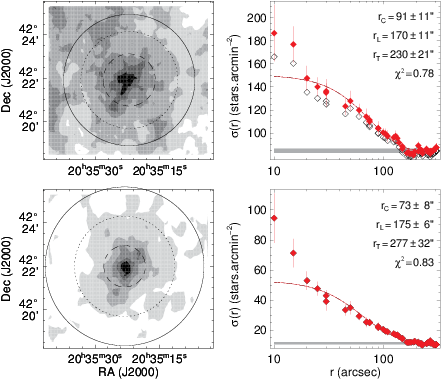}
\caption{Local stellar density maps (left panels) and radial density profiles (right panels) of LK12, sampling
stars with $K_s$$<$18.5 (upper panel) and $K_s$$<$15.5 (lower panel). The dashed, dotted and 
solid lines shown in the left panels represent the core, limiting and tidal radii, derived by the King-profile 
fitting (solid line) shown in the right panels. Completeness corrected sample (filled symbols with error bars) 
are compared to the raw sample (empty symbols).}
\label{fig:kpf}
\end{figure}

Fig.~\ref{fig:kpf} compares the result of the King-profile fitting for object LK12, using a completeness corrected 
fainter sample (K$_s$ $\le$ 18.5) and a brighter stellar sample (K$_s$ $\le$ 15.5), for which incompleteness 
is negligible. The derived core radius, tidal radius and limiting radius (i.e. the radius at which its density 
profile merges with the local background) are indicated on the density maps. It can be seen that the 
completeness correction is much more prominent on the fainter sample, making the distribution more peaked 
relative to the uncorrected sample. Furthermore, the RDP derived from the brighter stellar sample shows a 
reduced core radius when compared to the RDP derived from the full sample, indicating the high-mass stars 
are more centrally concentrated than the overall population, even when accounting for the higher incompleteness 
of the fainter population. This behaviour was also observed in the RDP's of BDB58 and Bica3, pointing out to a 
mass segregated population in these objects.

Additionally, FSR249 and LK15 presented a sub-clustered structure, which could not be described 
by King-like profile. However, they showed a coeval population that could be easily discerned from the field 
in the following CMD analysis and were therefore classified as true clusters. Their peak stellar densities 
were estimated from the RDP while their sizes were found by visually inspecting WIRCam images, Spitzer 
images and the star density maps. On the other hand, FSR245 and BDB55 presented irregular structures
that were found to be field populations. For all other clusters, their derived tidal radius was used to define 
the cluster size.

Although source crowding was effectively quantified and corrected in these clusters, extinction 
heterogeneity, particularly in the central regions of BDB57 and BDB59 could bias their derived 
structural parameters. In this case, we expect the stellar densities derived represent lower limits, 
while their core radius are actually upper limits. Table~\ref{tab:par} contains the mean stellar density 
($\bar{\rho}$), core radius (r$_c$) and tidal radius (r$_t$) for each cluster.

\subsection{Membership}

Like the majority of SFRs in the galaxy, the high extinction and relatively large distances of 
the stellar populations in Cygnus X makes proper motion data scarce or unavailable, impeding a 
kinematical derivation of membership for its star clusters. In this sense, photometric based membership has 
proven to be a fair substitute employing a multi-colour statistical approach, often coupled to positional 
arguments, to characterise and separate field and cluster populations \citep[e.g.][]{bb07, fssl10, dmco12, 
km14, pvp15}. In fact, even when kinematical data is available, recent works on star cluster characterisation 
have been including both positional data and multi-colour photometry in a multi-dimensional analysis to 
better constraint their derived memberships \citep[e.g.][]{sa09, s14}.

In our membership analysis we adopted the photometric method devised in \citet{maia10}, which assigns  
pertinence values to stars based on the local stellar density difference between a contaminated cluster 
sample and a control field sample in a multi-colour space according to the relation
\begin{equation} 
P_{phot} = \frac{\rho_{clu}-\rho_{fld}}{\rho_{clu}},
\label{eq:memb}
\end{equation}

\noindent
where $\rho_{clu}$ and $\rho_{fld}$ represents the local stellar density in the cluster and field samples, 
respectively, taken at the same position in the multicolour space. The method can also benefit from a 1-D 
positional argument representing the distance to the cluster centre to weight its assigned membership 
probabilities and better remove contaminants. The process is robust to the field sample selection as long as 
it is representative of the contaminating population in the cluster region. Generally, we have adopted the 
tidal radius as a limit to the cluster population and an annular ring around it with the same area as the 
inner region as the field sample, except when an inhomogeneity (e.g. another cluster or a 
dense cloud) is present. In such cases a manual selection of cluster or field region is often preferred.

Once a cluster and a field sample are supplied, a member list is readily defined from the derived 
membership probabilities by adopting a suitable cutoff threshold. Previous work with the method with a 
large and diverse cluster sample \citep{maia14} has shown that a cutoff at 30\% to 50\%, depending on 
the level of field contamination, is optimal to clearly reveal cluster sequences in the CMD.

\begin{figure}
\centering
\includegraphics[angle=90,width=\linewidth]{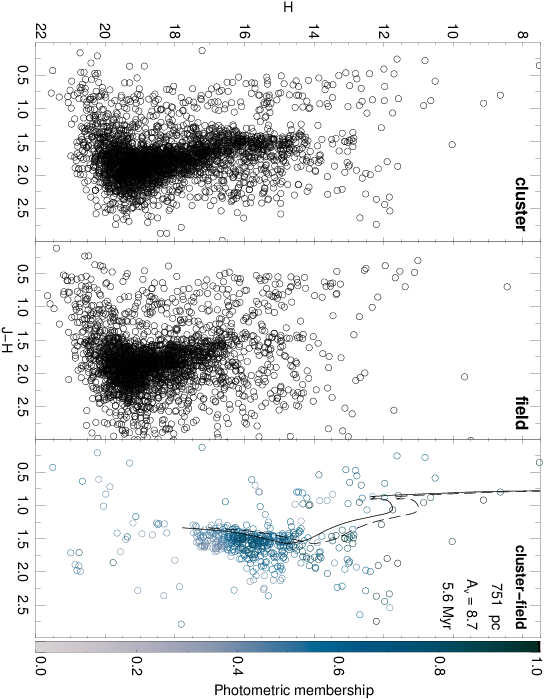}
\caption{CMD comparing the samples in LK12 region (left panel), in its nearby field (middle panel) and 
the field-subtracted sample (right panel). The colour bar shows the assigned photometric probabilities. 
Isochrones representing the single star (solid line) and equal mass unresolved binaries (dashed line) 
populations are shown, along with the most probable age, distance and extinction inferred.}
\label{fig:memb}
\end{figure}

Fig.~\ref{fig:memb} shows the result of such procedure comparing the cluster, field and the decontaminated 
sample of LK12, projected in the $H$ vs $J$-$H$ CMD. Although some outlying contaminants are still 
present, the method has been able to successfully separate most of the visible cluster sequence. 

\subsection{Extinction}

Since most SFR present large quantities of the parental cloud gas and dust, modelling extinction has 
become a crucial step in studying these regions, particularly through photometric data. In this sense, 
near-infrared data has been playing a crucial role as it can reach significantly deep in these gas rich 
environments and most importantly due to its availability through several ground-based surveys (e.g. 
2MASS, VISTA) and observing facilities.

To this end, far-optical to mid-infrared colour excesses have been used in conjunction to known extinction 
laws to derive extinction maps in SFR spanning a diversity of spatial scales \citep[e.g.][]{lad06, s11, 
mzn11}. While this approach may underestimate extinction in the most obscured areas (due to the 
absence of stars beyond the obscuring cloud), the derived values are often well correlated to CO maps, 
which are regarded as good tracer of cold, dense clouds and does not require additional data (HI, dust 
emission) other than the stars photometry.

Since our sample is expected to be composed by a coeval population of pre main sequence low mass stars, 
down to $\sim$0.1 M$_\odot$, we have refrained from adopting a fixed intrinsic colour value or relying on
calibrated intrinsic colours for our extinction calculations, since these are more appropriate for dwarf stars 
of later spectral types. Instead we have opted to use the intrinsic colours derived from stellar evolutionary 
models for solar metallicity stars.

\begin{figure}
\centering
\includegraphics[angle=90,width=\linewidth]{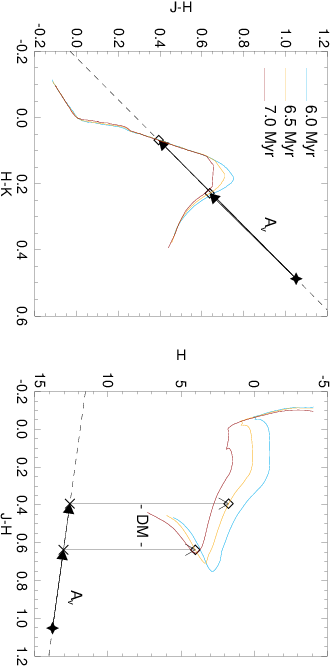}
\caption{Colour-colour (left) and colour-magnitude (right) diagrams comparing solar metallicity isochrones 
of $\log$t 6.0, 6.5 and 7.0. It shows how the reddening vector (dashed lines) can be used to derive both 
the extinction and distance modulus for a particular star, given its possible isochrone intercepts (diamonds).}
\label{fig:ext}
\end{figure}

Although many models are currently available in the literature, we have found that very few provide 
tracks for pre-main-sequence stars (PMS) covering stellar masses from $\sim$0.1 to possibly 10-50 
M$_\odot$, namely the model by \citet[][hereafter SDF00]{sdf00} reaching 7 M$_\odot$ and the model 
by \citet[][hereafter PARSEC]{b12}, reaching over 300 M$_\odot$. Fig.~\ref{fig:ext} compares the 
PARSEC isochrones of ages from 1 to 10 Myr in the $H$ x $J$-$H$ and $J$-$H$ x $H$-$K_s$ diagrams. 
It can be seen from the colour-colour diagram that for such young ages, that the overall colour excess 
differences arising from such large age uncertainties ($\Delta \log$t = 1.0 dex) would be very small possibly 
reaching $\Delta$($J$-$H$)$_0\sim0.1$ in certain mass domains. In addition, the CMD shows that  
different aged PMS populations could theoretically be discerned, even in the absence of giant stars, due 
to the increasing colour extent between the main-sequence turn in point and the reddest turn around 
$\sim$0.5 M$_\odot$ ($J$-$H$ $\sim$ 0.7), possibly tied to the developing of a radiative core. 
However, these differences would require a colour resolution inferior to $\sim$ 0.2 mag at the relatively 
faint red turn.

Given a particular model isochrone, we have used the extinction law of \citet{rk85} to constrain the slope
of the reddening vector in the colour-colour diagram and find the isochrone intercepts that define
the intrinsic colours of a particular star. As shown in Fig.~\ref{fig:ext}, most of our stars will intercept the 
isochrones at two different points, corresponding to a stellar mass smaller/greater than $\sim$ 0.7 
M$_\odot$. Afterwards, extinction estimative to each of these intercepts can be easily calculated through 
one of the following relations:
\begin{align}
\mathrm{A}_v^{jh} =&\ 9.725 \times \mathrm{E}\,(J-H) \label{eq:avj} \\
\mathrm{A}_v^{hk} =&\ 15.314 \times \mathrm{E}\,(H-K_s) \label{eq:avh}  \\
\mathrm{A}_v^{jk} =&\ 5.948 \times \mathrm{E}\,(J-K_s) \label{eq:avk}
\end{align}

For most stars these equations return identical values, as expected.
However, we have opted to use the extinction derived from $J$ and $H$ bands (first equation) as they 
are less likely to be affected by infrared colour excess, shown more strongly in $K_s$ band. Moreover, 
these bands provide a larger colour difference between the lower and higher mass isochrone intercepts 
(see Fig.~\ref{fig:ext}), which makes easier the determination of the right intercept.

As a byproduct of using isochrones for determining the intrinsic colours, distance modulus for each intercept 
can also be determined through one of the following equations:
\begin{align}
\mathrm{DM}^j =&\ J- J_0 - 0.28765 \times A_v \label{eq:dmj} \\
\mathrm{DM}^h =&\ H - H_0 - 0.18482 \times A_v \label{eq:dmh} \\
\mathrm{DM}^k =&\ K_s - K_0 - 0.11952 \times A_v \label{eq:dmk}
\end{align}

As with the extinction, these equations returns very close values for most of stars, save for a few presenting 
$K_s$ band excess. To avoid these cases we have adopted the median of such values for improved 
accuracy. 

Extinction uncertainties were evaluated for by properly propagating the photometric errors through 
equations (\ref{eq:avj}-\ref{eq:avk}). Likewise, distance modulus uncertainties were calculated by taking into 
account both the photometric and extinction uncertainties for a star. Note that these errors are invariant, 
whatever choice is made regarding the isochrone intercepts of a particular star.

Since a smooth extinction solution is often expected through these young populations, determination of 
the 'best' isochrone intercept for each star can be done through a global minimisation of the extinction 
dispersion. This was done in a two-step process in which all stars were initially assigned to its lowest 
possible extinction values (corresponding to lower mass isochrone intercept) and later iterated through 
their possible extinction values to find the configuration that minimises the dispersion of the whole sample.
Note that although this configuration also constrains a 'best' distance modulus for each star, 
it usually does not minimizes the dispersion of the distance moduli.

Once such configuration was found an extinction probability density function (PDF) was built for each star 
using its 'best' inferred extinction value and assuming a normal distribution of its uncertainty. A global 
extinction distribution was then created by stacking the individual extinction PDF of each star. The 
expected extinction value for the population was finally estimated as the peak value of this global extinction
distribution. Similarly, a global distance modulus distribution was also derived and an expected distance 
modulus value evaluated from its peak. 

\begin{figure}
\centering
\includegraphics[width=\linewidth]{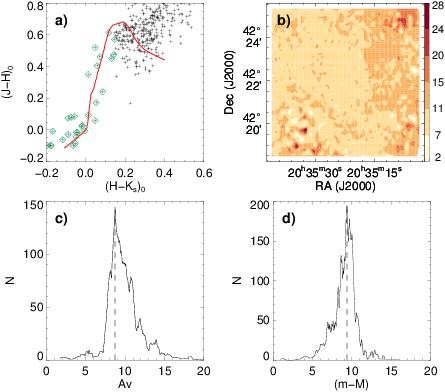}
\caption{Results of the extinction determination showing (a) the adopted isochrone (solid line) and the 
intrinsic colours of LK12 stars, dereddend using the expected extinction value for the population. Most stars 
(plus signs) were assigned to their lower mass intercept in the isochrone while a few bluer ones (diamonds) 
where assigned to the higher mass intercept. The derived extinction values were interpolated into a 
local extinction map (b), with the colour bar indicating the extinction levels, and used to construct the
global extinction (c) and distance modulus distributions (d) for LK12 stars from which their expected values 
(dashed lines) were derived. }
\label{fig:ext2}
\end{figure}

Fig.~\ref{fig:ext2} shows the result of such procedure over the stars in the vicinity of LK12. 
It can be seen that the inferred 'best' isochrone intercepts have successfully matched most of the stars 
to the lower mass domain of the isochrone in the colour-colour diagram, while assigning slightly bluer 
stars to the higher mass interval, providing a good dereddening solution for the cluster population. 
An extinction map was promptly derived from the full local stellar sample by interpolating their extinction 
values into an uniform grid with a desired (5\arcsec-15\arcsec) resolution. A$_v$ and DM distributions, in 
the other hand, are best derived from a field-cleaned sample, in order to increase the statistical significance 
of the cluster population and improve the determination of their expected values. 

Although small fluctuations can be seen through the global extinction distribution, it is clear that the 
population shows a very pronounced peak value with positively skewed tail. This means that 
the expected extinction value actually represents the minimum extinction towards the population, or the 
foreground extinction shared by a larger fraction of stars. Intra cluster extinction is often variable and 
therefore won't form secondary peak in the distribution, but will produce the observed tail towards larger 
extinction values.

\subsection{Distance and age}

Even though extinction can be well constrained through the stars colours, determination of distance and 
age of a stellar population through photometry is met with several difficulties as many phenomena can 
displace the stars from their fiducial position in the CMD. Intra-cluster nebulosity and variability 
(which may arise intrinsically or due to surrounding material) are likely the most unaccounted 
effects. These are, however, somewhat attenuated on populous objects as they are likely to be better 
sampled across the mass spectrum and to posses a greater population of stars in their 'right' position 
in the CMD from which the parameters can be more reliably determined. 

In this context, several methods have been developed recently to reliably determine ages from 
CMD data using a variety of approaches such as statistical indicators, bayesian inference and synthetic 
populations \citep[e.g.][]{jl05, nj06, mdc10}. Although many of these methods can lift the known 
degeneracies between some isochrone parameters (e.g. age-metallicity, reddening-age), they are often 
not optimised to deal with pre-main-sequence populations showing very heterogeneous stellar
distribution over the CMDs.

In our approach, although neither the derived extinction distribution nor its expected value are expected to 
change considerably with the adoption of different aged isochrones, it is very likely that the distance modulus 
distribution will (see Fig.~\ref{fig:ext}). In order to determine which isochrone best describes a particular 
population we have adopted a strategy very similar to the one used for deriving extinction, seeking to minimize
the dispersion of distance modulus distribution across a range of ages. This was quantified by defining a 
coefficient to be maximised, that measures how sharp the extinction and distance modulus distributions 
derived from a particular isochrone are. It was defined as:
\begin{equation}
S\,(\log t) = \Big[ \frac{N_{max}}{W} \Big]_{Av} \times \Big[ \frac{N_{max}}{W} \Big]_{DM}
\end{equation}

\noindent
where $N_{max}$ represents the peak value and $W$ the FWHM of either the extinction ($Av$) or 
distance modulus ($DM$) distribution derived from a particular isochrone.

Such method provides an advantage over curve-fitting or synthetic CMD techniques in the sense that 
it is very robust to outliers such as bright field stars since they will show a range of 
unrelated extinction and distance modulus values that will not disturb the contained distribution of the
dominant stellar population. 

\begin{figure}
\centering
\includegraphics[width=\linewidth]{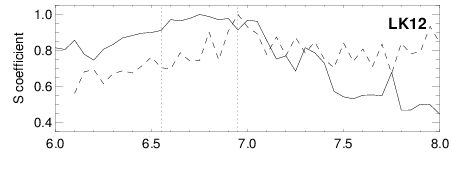} \vskip -0.25cm
\includegraphics[width=\linewidth]{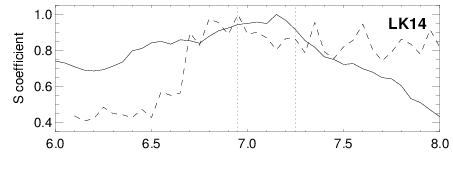} \vskip -0.25cm
\includegraphics[width=\linewidth]{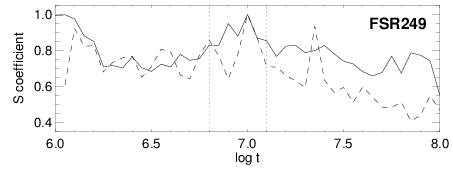}
\caption{Normalised $S$ coefficients derived from SDF (dashed lined) and PARSEC (solid line) isochrones 
spanning 1-100 Myr for objects LK12, LK14 and FSR249. The most probable age interval adopted for each 
object is indicated by the vertical dotted lines.}
\label{fig:age}
\end{figure}

The procedure described on the last section was then repeated for the full range of solar metallicity 
PARSEC isochrones aged from 1 Myr to 1 Gyr and the corresponding $S$ coefficient determined for each 
one of them. This was also done using the solar metallicity SDF isochrones from 1 Myr to 100 Myr, taken 
at the exact same ages as the former ones. Fig.~\ref{fig:age} compares the normalised $S$ coefficients 
as a function of age for the first three objects of our list. It can be seen that both isochrone models provides 
$S$ peaks at compatible ages, within a 0.2 dex margin, with the PARSEC tracks showing a smoother 
solution, possibly due to its larger mass coverage and finer sampling across the mass spectrum. 

A most probable age interval was then defined as the region around the $S$ peak, taken from the PARSEC 
isochrones, where the $S$ values showed an enhanced value over the local 'pseudo-continuum' distribution
(see Fig.~\ref{fig:age}). For our objects this interval spanned an age range inferior to 0.4 dex, which most 
of the time corresponds to an uncertainty of 0.2 dex around a central value. This uncertainty is about the 
same order as the one arising from the adoption of different isochrone models. Combining these values 
would result in a general age uncertainty of $\sigma_{\log t}$ = 0.28 dex. 

The isochrones corresponding to the derived age interval are likely to produce very similar extinction 
distributions while producing distance modulus distributions peaked at different values.  
As Fig.~\ref{fig:ext} shows, the youngest isochrone is likely constrain the maximum distance 
modulus value, with the oldest one constraining the minimum value. Therefore, we have used the range 
of expected distance moduli derived from these isochrones to determine a most probable distance interval
for each object. For LK12, the cluster presenting the largest probable $\log t$ range in our sample 
($6.55 \le \langle \log t \rangle \le 6.95$), the expected (peak) distance moduli have shown values between 
9.08-9.58 mag while the expected (peak) extinction values varied between 8.69-8.82. This would translate 
into a most probable distance interval between 655-825 pc and into an extinction uncertainty of 
only $\sigma_{Av} = 0.07$ mag. 

The derived most probable values of age, distance and extinction were used to superimpose the 
single-star sequence and that corresponding to equal-mass unresolved binaries over the CMD of LK12, 
shown in Fig.~\ref{fig:memb}. The model sequences appears to be in good agreement with the loci of 
the cluster stars.

\subsection{Mass distribution}

There is ongoing debate of whether star formation follows a universal mass distribution - 
described by the initial mass function (IMF) \citep[e.g.][]{kr13,c05} - or if it can be affected by the 
properties of the local environment. Although there is growing evidence showing that the IMF appears to be
constant across different SFRs and even in the sub-stellar regime \citep{bcm10,o14}, 
the matter is not settled \citep[e.g.][]{t11,dib14}.

In this sense, the distribution of stellar masses in open clusters is probably the most used method to 
investigate the IMF due to the advantages in studying a coeval population: uniform distances, ages and 
reddening. Embedded young clusters provide even more appealing subjects, as they allow sampling of 
more massive stars and due to the fact that their mass distribution may have not had the time to 
dynamically process or evolve, thus being closer to its true initial conditions.

Derivation of stellar IMF is usually difficult, especially in young populations. Apart from the usual problems in 
dealing with obscured regions, it also requires knowledge of mass-luminosity relationships 
appropriate for pre-main-sequence phase, which are currently scant. Since we are employing PMS stellar 
models to characterise our stellar populations, once a particular model has been found to best reproduce an 
observed sample, the stellar masses can be derived from the isochrone mass-luminosity relationship. 

\begin{figure}
\centering
\includegraphics[width=\linewidth]{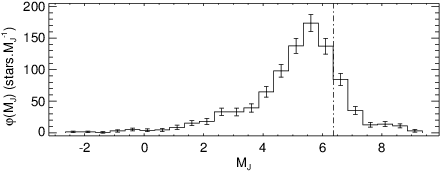} \vskip 0.2cm
\includegraphics[width=\linewidth]{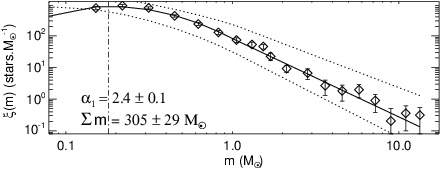}
\caption{Completeness corrected $M_J$ luminosity function (top) and corresponding mass function (bottom) 
of LK12. Error bars indicate the monte-carlo derived uncertainty (see text). The vertical dot-dashed lines 
indicate the fainter limit of the mass function fitting range. The solid line indicates the best fit to a log-normal 
mass-function while the dotted ones delimit the reference IMF domain. The derived total mass and the fitted 
high-mass exponent are shown.}
\label{fig:mass}
\end{figure}

In order to achieve that, the star's apparent magnitudes were converted to their absolute values using 
the most probable extinction and distance modulus derived for the population. Furthermore, the 
uncertainties in these parameters (see previous section) were properly added to the photometric 
errors to calculate realistic uncertainties for the absolute magnitudes.
Assuming a normal distribution of these uncertainties, a monte-carlo procedure was then employed to
construct a binned $M_J$ luminosity function which was afterwards corrected for the local incompleteness, 
as derived from our artificial star simulations (see Fig.~\ref{fig:complcor}). 
The isochrone mass-$M_J$ relationship was then used to convert this completeness corrected luminosity 
function and its monte-carlo derived uncertainty to a mass distribution, with stars fainter/brighter than the 
isochrone mass limits being discarded.
Again, PARSEC isochrones were preferred due to their largest mass coverage ($0.1 \le m (M_\odot) \le 350$). 
Fig.~\ref{fig:mass} shows the luminosity and mass functions derived from the assigned members of LK12. 

A log-normal analytical IMF function from \citet{c05}, hereafter refereed as the IMF, was fitted to the 
computed mass function to derive a characteristic mass ($m_c$), mass dispersion ($\sigma_m$) and 
high-mass power-law exponent ($\alpha$) for the distribution, through the following relations:
\begin{equation} \label{eq:mf}
\begin{split}
\xi (m)  = \frac{dN}{dm} &= \frac{A}{m}\,\mathrm{exp}\left[- \frac{(\log m - \log m_c)^2}{2\,\sigma_{m}^2} \right], 
& m \leq 1\,\mathrm{M}_\odot \\
& = A\,k\,m^{-\alpha}, & m \geq 1\,\mathrm{M}_\odot
\end{split}
\end{equation}

\noindent
where $A$ and $k$ are normalisation constants. The fit was limited in the fainter domain by either the 95\% 
completeness limit or by faintest luminosity bin where the cluster sequence could be clearly identified in the CMD.

We have also adopted its parametrisation of the galactic field 
system IMF as reference values: $m_c$ = 0.25$\pm$0.05, $\sigma_m$ = 0.55$\pm$10 and 
$\alpha$=2.35$\pm$0.36, where the uncertainty of the first two parameters were arbitrarily set to 20\% 
level, due to their absence. These reference values and their uncertainties were used, along with the 
normalisation (and its uncertainty) derived for each cluster, to define a domain in the mass distribution 
plot inside which stars are said to comply with the IMF. This was meant to provide an uncertainty derived 
scatter around the nominal IMF parameters. Both the fitted function and the derived IMF compliance domain 
are shown in Fig.~\ref{fig:mass}.

Total stellar mass and its uncertainty was also estimated by summing contributions from individual mass 
bins. This was also done by assuming a population composed entirely of equal mass unresolved binaries 
(i.e. shifting the derived luminosity function by 0.75 mag and doubling the star counts). Since the difference 
between such estimates were larger than the $\sim$10\% uncertainty derived in each case, we have 
instead reported the total cluster mass as a range were the lower limit corresponds to the mass 
derived from the single star (unmodified) population and the upper limit corresponds to the mass derived
for a population of binaries. The true cluster mass is likely to be in between. 

In addition, whenever a region smaller than the derived cluster size had to be used in the analysis (i.e. due 
to a nearby feature or object), a correction factor was calculated from the ratio of the expected number of 
stars inside this region to the expected total number of stars inside the cluster tidal radius, and applied to 
to derived total mass. These expected numbers of stars were obtained by integrating the cluster 
analytical King profile function through the appropriate domains. For example, we have restricted the analysis 
of LK12 to its inner 173\arcsec due to nearby nebulosity, even though its computed tidal radius was 
252\arcsec. Integration of its King profile in the inner region yields 454 stars, comparable to the 
decontaminated sample of 449 stars, used in the analysis. Integration up to its tidal radius yields 473 stars, 
implying that about 96$\%$ of its total stellar population was sampled in the inner region. Therefore, a
correction factor of 1/0.96 was applied to its derived mass to compensate for the unaccounted population in 
the outer region of the cluster.

\section{Individual results and comments}

The methodology described in Sect.~\ref{method} was employed to study the catalogued objects in the 
region. Previous literature references and potential comments on the analysis of each object are given in 
this section. 
For most of the clusters image cuts from our NIR mosaics are compared to Spitzer image
cuts taken from the processed mosaics of the {\it Cygnus-X Spitzer Legacy Survey}\footnote{
https://www.cfa.harvard.edu/cygnusX/} downloaded from NASA/IPAC Infrared Science Archive (IRSA).
A compilation of the most important physical parameters derived is shown in Table~\ref{tab:par}, including
the mean stellar density ($\bar{\rho}$), the core (r$_c$) and tidal (r$_t$) radii, the mean extinction (A$_v$)
and the mass function characteristic mass (m$_c$), mass dispersion ($\sigma_m$) and high mass slope 
($\alpha$). Objects FSR245, BDB55 and BDB60 are not shown, as we have found the former two to not 
represent physical systems and could not reliably determine the parameters of the latter.

\begin{table*}
\caption{Derived physical parameters for clusters and candidates}
\setlength{\tabcolsep}{3pt}
\begin{tabular}{lcc r@{$\pm$}l r@{$\pm$}l r@{$\pm$}l c r@{$\pm$}l c r@{-}l r@{$\pm$}l r@{$\pm$}l r@{$\pm$}l }
\hline
ID & RA & Dec & \multicolumn{2}{c}{$\bar{\rho}$} & \multicolumn{2}{c}{r$_c$} & \multicolumn{2}{c}{r$_t$} & 
Age & \multicolumn{2}{c}{A$_v$} & D & \multicolumn{2}{c}{Mass} & \multicolumn{2}{c}{m$_c$} & 
\multicolumn{2}{c}{$\sigma_{m}$} & \multicolumn{2}{c}{$\alpha$} \\ 
& ($^h$:$^m$:$^s$) & ($^\circ$:\arcmin:\arcsec) & \multicolumn{2}{c}{(pc$^{-2}$)} & 
\multicolumn{2}{c}{(pc)} & \multicolumn{2}{c}{(pc)} & (Myr) & \multicolumn{2}{c}{ } & (kpc) & 
\multicolumn{2}{c}{(M$_\odot$)} & \multicolumn{2}{c}{(M$_\odot$)} & \multicolumn{2}{c}{($\log$\,M$_\odot$)} \\ 
\hline
  LK12 & 20:35:22 & 42:21:54 & 174& 18 & 0.29&0.02 & 0.92&0.07 & 4-9  &  8.73&0.07 & 0.65-0.83 & 318&441 & 0.34&0.04 & 0.34&0.02 & 2.44&0.10 \\
  LK14 & 20:35:42 & 42:29:31 & 274& 40 & 0.23&0.03 & 0.84&0.10 & 9-18 &  8.49&0.04 & 0.92-1.12 & 433&621 & 0.38&0.13 & 0.30&0.04 & 2.08&0.10 \\
FSR249 & 20:38:29 & 42:29:15 & 100& 39 & 0.44&0.19 & 0.93&0.33 & 7-13 &  7.00&0.12 & 1.45-1.75 & 312&458 & 0.38&0.19 & 0.39&0.08 & 1.83&0.11 \\
  LK13 & 20:38:29 & 42:06:22 & 922&101 & 0.08&0.01 & 0.25&0.01 & 4-6  &  7.19&0.06 & 0.45-0.55 &  38& 46 & 0.20&0.06 & 0.35&0.03 & 1.98&0.34 \\
 BDB59 & 20:38:37 & 42:37:29 &  77& 17 & 0.18&0.03 & 0.72&0.11 & 1-3  & 12.95&0.03 & 1.15-1.38 & 224&274 & 0.22&0.12 & 0.50&0.09 & 1.41&0.32 \\
 BDB58 & 20:38:54 & 42:22:44 &  46&  6 & 0.40&0.10 & 1.30&0.10 & 1-3  & 13.47&0.06 & 2.74-3.34 & 610&772 & 0.30&0.10 & 0.37&0.07 & 1.53&0.29 \\ 
 BDB57 & 20:39:02 & 42:19:37 & 117& 63 & 0.04&0.01 & 0.40&0.15 & 1-3  & 24.40&0.06 & 0.60-0.78 & 163&218 & 0.18&0.10 & 0.46&0.07 & 1.36&0.12 \\
 Bica3 & 20:40:01 & 42:18:17 &  12&  1 & 1.34&0.11 & 3.85&0.25 &$\sim$500& 15.10&0.03 & 4.49-5.10 & \multicolumn{2}{c}{$\sim$3000} & \multicolumn{2}{c}{---} & \multicolumn{2}{c}{---} & 1.33&0.14 \\
  LK15 & 20:40:30 & 42:03:13 &  70& 22 & 0.44&0.16 & 1.02&0.28 & 1-3  & 11.18&0.18 & 1.65-2.13 & 367&475 & 0.55&0.26 & 0.33&0.11 & 1.67&0.10 \\
 BDB56 & 20:41:19 & 41:58:09 &  97& 22 & 0.15&0.01 & 0.58&0.06 & 2-5  &  9.80&0.13 & 1.39-1.61 & 140&186 & 0.41&0.18 & 0.40&0.08 & 1.79&0.11 \\
 \hline
\end{tabular} 
\setlength{\tabcolsep}{6pt}
\label{tab:par}
\end{table*}

Although some of our targets have been previously processed by \citet[][hereafter K13]{k13} pipeline we 
have found that many of their derived parameters are at odds with the ones determined in this work. 
Because proper motion in the Cygnus-X region is mostly unavailable or unreliable, their analysis relied 
heavily on 2MASS photometric data, often extracted in regions many times greater than the clusters 
projected size with scant numbers of probable members. Furthermore, it appears that for the majority of 
young clusters (those lacking a turn-off), age could only be estimated by eye fitting of an isochrone. 
Therefore, although comparison with their results are provided, we found that they might be unreliable for 
young, very reddened and/or compact objects, like many of our targets.

\subsection{LK12}

LK12 is frequently referred as DR17 IR cluster, given its proximity to this bright radio source and HII region.
The compact X-ray source AXJ2035.4+4222-Src2 lying on the centre of the cluster was found to be 
probably associated with Cygnus Rift at $d \leq 1.0$ kpc \citep{rrk01}, being the strongest constraint
to the cluster's youth and distance. Although two spectroscopic confirmed OB stars could be found in the 
cluster periphery, their derived extinctions differ significantly from the cluster mean value, suggesting that 
they might belong to the field. Indeed, one of them (TYC 3159-6-1) was reported as a runaway B1 
supergiant with A$_v$=5.9, age of 4.0-4.5 Myr and distance between 1.2 and 1.5 kpc \citep{g14}, while the
other (TYC 3161-106-1) was reported as a Cygnus OB2 field B1III star with an age of 14 Myr, A$_v$=3.7 
and distance of 1.4 kpc \citep{cp12}.

\begin{figure}
\centering
\includegraphics[width=\linewidth]{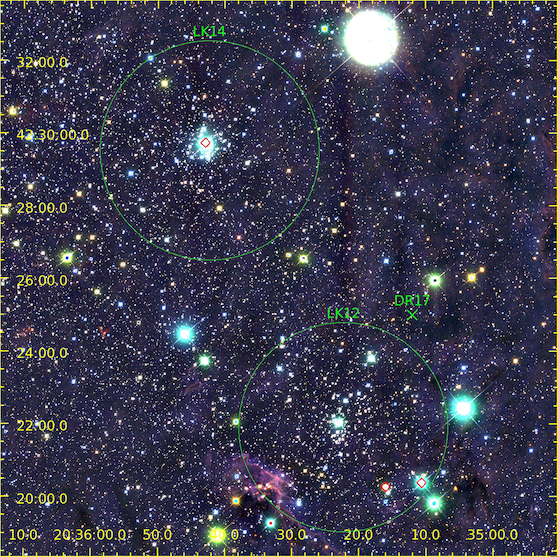} \\ \vspace{3mm}
\includegraphics[width=\linewidth]{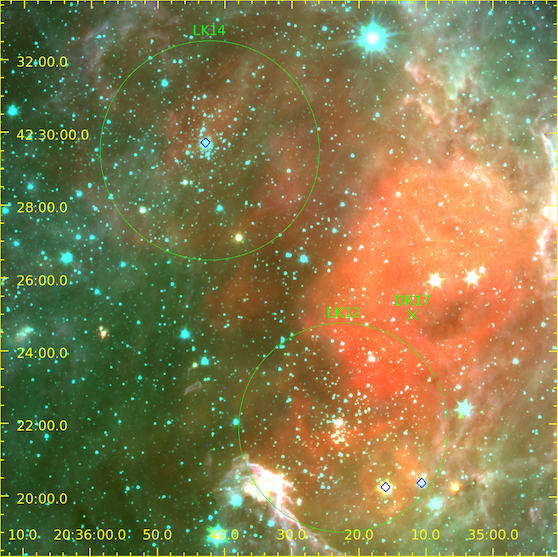}
\caption{Colour-composite images of the region around DR17, as seen from our $J$, $H$ and 
$K_s$ images (top) and from Spitzer $[3.6]$, $[4.5]$ and $[24] \umu m$ channels (bottom). The extraction 
radius for LK12 and LK14 are indicated, along with nearby, known B-type stars (diamonds).}
\label{fig:lk12}
\end{figure}

The Spitzer composite image shown in Fig.~\ref{fig:lk12} reveals the presence of a pillar-like structure 
beyond the southeast border of this cluster that appears to point north, towards LK14, while being ionised 
from the northwest. Although LK12 lies closer to the pillar, the direction of nearby cometary tails reveals that 
the region is being ionised by the brighter stars at the center of DR17, approximately 5\arcmin northwest.
CO emission velocity channel maps by \citet{s06} seems to suggest that W75N and DR17 are somewhat 
connected, as their emission peaks at the same velocity channels. This was also found by G12 which 
assigned a distance of 1.0-1.8 kpc to both regions. 

\citet[hereafter, DB01][]{db01} were the first to identify this cluster assigning a visual extinction of 15 and a 
distance of 1.2 kpc. It was later revisited by LK02, who reported a visual extinction between 
9-18 and a distance between 1.3-2.0 kpc. Furthermore, an integrated mass estimative of 
600-1760 M$_\odot$ was calculated by extrapolating a fitted high mass slope of $\alpha$=2.61$\pm$0.25.
Both works have used 2MASS data and relied on general M/L relations to derive the cluster physical 
parameters rather than directly determining an age for the cluster. This cluster was also processed by K13, 
reporting an age of 4 Myr, A$_v$=10 and distance of 1.7 kpc. 

Our results point out that LK12 is a young (4-9 Myr) emerging (A$_v$$\sim$9) cluster and that it lies 
closer than reported by previous studies, at a distance of 650-825 pc. These parameters would imply a 
diameter of 1.8 pc, and a stellar mass distribution that appear to be consistent with the one by LK02 in the 
high-mass range and that follow the IMF down to 0.2 M$_\odot$, totalling 315-440 M$_\odot$. 
Because of a heavy extinction column in its eastern region, the analysis of this
cluster was based in a reduced sample, extracted at a inner radius (173\arcsec) rather than at its tidal radius 
(252\arcsec). Integration of its derived King-profile reveals that this reduced sample corresponds to 
$\sim$96\% of the sample expected at its tidal radius. Therefore, the reported total mass estimative has
been corrected by this factor, assuming a homogeneous mass distribution. The main charts involved in 
this cluster analysis are compiled in Fig.~\ref{fig:ap_lk12}.

\subsection{LK14}

LK14 is located about 8\arcmin north of LK12 and could also be related to D17 SFR, given their proximity. 
Its stellar population, as seen in Fig.~\ref{fig:lk12}, appears considerably more evolved than LK12, 
presenting less extinction and being composed of fainter stars with a more centrally condensed distribution 
than its neighbour. Its central star has been spectroscopically observed by \citet{raw00} and classified as a 
O7-B9.5 star with A$_v$=9.2$\pm$0.3. 

This cluster was first identified by LK02 , acknowledging it as an evolved, nearby object with 
A$_v$ $\sim$ 8.5. They derived a tentative spectral type for the bright central star, obtaining B0 or O5.5 
whether the assumed distance was about 800 or 1250 pc, finally adopting the latter result and rejecting the 
central star as a foreground object. An integrated mass estimative of 470-1160 M$_\odot$was also calculated 
by extrapolating a fitted high mass slope of $\alpha$=2.51$\pm$0.35.

We have found this cluster to be significantly older than its neighbour presenting an age between 9-18 Myr, 
A$_v$=8.5 and distance between 920-1120 pc. These parameters would imply a 1.7 diameter cluster with a 
stellar mass distribution that is consistent with the one by LK02 in the high-mass range and that appears to 
follow the IMF down to 0.3 M$_\odot$, totaling 430-620 M$_\odot$.
Concerning the brightest star in the center of the cluster, we have used the \citet{raw00} optical $B$ and $V$ 
magnitudes for this star along with \citet{co06} stellar calibrations and our derived distance modulus and 
extinction to estimate an spectral type B0-B1 for this star. This is in accordance with the mass estimate of 
$\sim$14 M$_\odot$ given by the adopted isochrone models for our $JHK_s$ photometry. Since that this 
star also shows a extinction that is very consistent with the cluster mean value, we believe that it does 
belong in the cluster and might have held an important role in shaping the local environment in DR17. The main 
charts involved in this cluster analysis are compiled in Fig.~\ref{fig:ap_lk14}.

\subsection{FSR245}

\begin{figure}
\centering
\includegraphics[width=\linewidth]{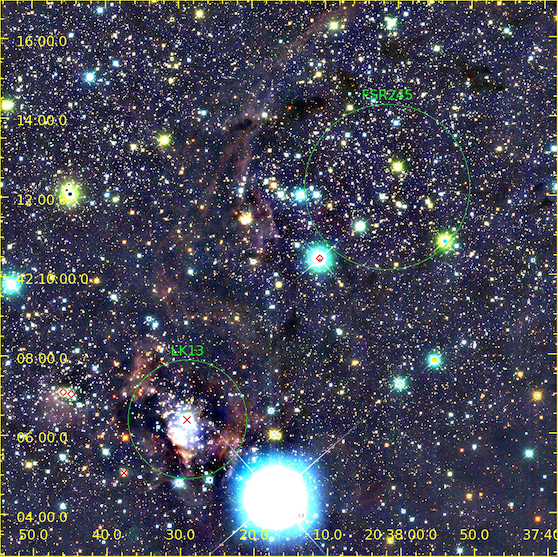} \\ \vspace{3mm}
\includegraphics[width=\linewidth]{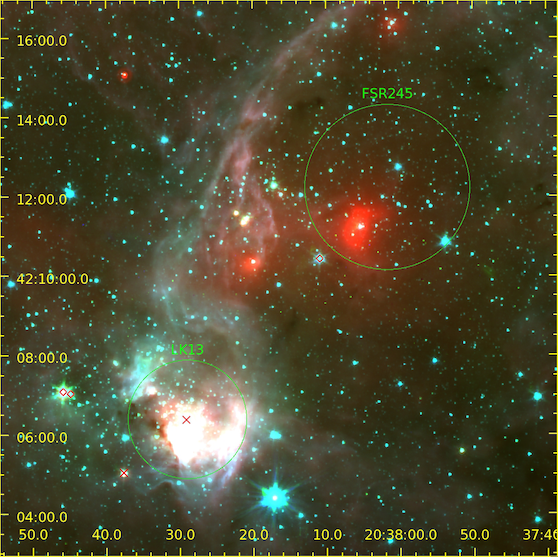}
\caption{Colour-composite images of the region around the "diamond ring", as seen from our $J$, $H$ and 
$K_s$ images (top) and from Spitzer $[3.6]$, $[4.5]$ and $[24] \umu m$ channels (bottom). The extraction 
radius for FSR245 and LK13 are indicated, along with nearby, known B-type stars (diamonds) and K0 stars 
(crosses).}
\label{fig:lk13}
\end{figure}

This cluster candidate was identified by \citet{fsr07} by using stellar density counts from 2MASS 
catalogue. It was later included in one of the regions analysed by \citet{b10} around the "diamond ring" 
(see Sect.~\ref{sect.lk13}), reporting it to be relatively older than the surrounding regions. 
It lies close to the radio source W70, and features a bright blue star with considerable 24$\umu$m
emission that appears to have cleared out the gas and dust around it. Despite that, the region appears 
featureless both in our WIRCam and Spitzer images (Fig.~\ref{fig:lk13}), presenting a highly fragmented
stellar distribution that is not compatible with a King-profile. 

Our analysis has found two most probable age peaks centred around 10 Myr and 500 Myr for this object.
However, these peaks are much broader than found for the other analysed clusters indicating that they
do not represent coeval populations, representing instead two overlapped field populations with very 
different mean ages. The stellar extinction and distances also present bimodal distributions with an 
abundant, exposed (A$_v \sim$ 3) population at 1.2 kpc and a highly reddened (A$_v \sim$ 11) one 
at much greater distances, possibly related to background Galactic arms. Although they can be identified
on the CMD, none of these two populations show any spatial correlation, both being dispersed over the 
field.

We conclude that FSR245 is not a real cluster, but instead a region inside the "diamond ring" 
that has been cleared out of the local dust and gas content by a nearby massive star, 
thus allowing for the emergence of the background population of the Galactic disk. That is probably the 
cause of the local stellar density enhancement that led to the initial identification of this object.

\subsection{FSR249}

\begin{figure}
\centering
\includegraphics[width=\linewidth]{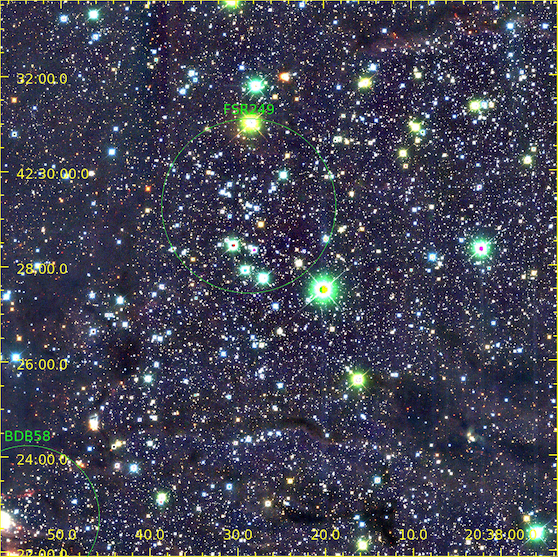} \\ \vspace{3mm}
\includegraphics[width=\linewidth]{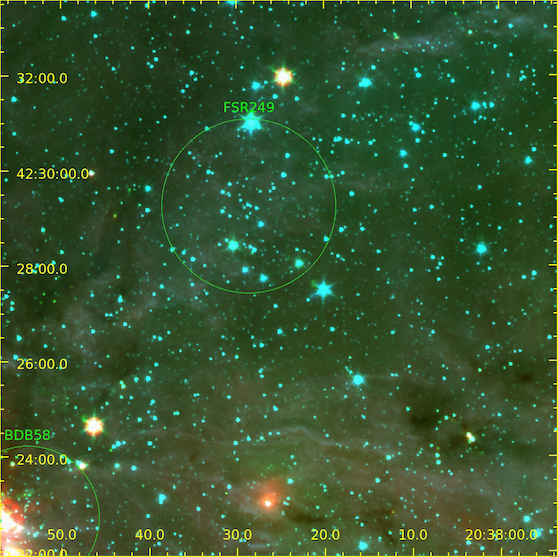}
\caption{Colour-composite images of the region between FSR249 and DR21, as seen from our $J$, $H$ and 
$K_s$ images (top) and from Spitzer $[3.6]$, $[4.5]$ and $[24] \umu m$ channels (bottom). The extraction 
radius for FSR249 is indicated in the figure.}
\label{fig:fsr249}
\end{figure}

This cluster candidate was identified by \citet{fsr07} and subsequently processed by K13, classifying it 
as a large (10 pc), young (1 Myr) and emerged (A$_v$ $\sim$ 4) cluster at 1.8 kpc. Fig.~\ref{fig:fsr249} 
shows that although it is hard to distinguish it from the rich background on our images, its brighter 
population is better constrained in Spitzer frames revealing it to be inside a gaseous filament extending 
irregularly to the northwest but converging towards DR21 SFR, about southeast. Star density maps 
of FSR249 shows that although it presents a dense central clump, it still does not have a globular-like 
shape and therefore converged very poorly to our King-profile fittings. Therefore, we have arbitrarily
defined a $\sim$2\arcmin radius (close to the derived tidal radius of 110\arcsec) based on the visual 
inspection of the stellar density maps and on the extent of the gaseous structure seen on its Spitzer image.

Our analysis have found two age ranges that are equally probable at 1-2 Myr and 7-13 Myr. Although the 
younger one appears to be in agreement with the age by K13, we believe that the older age range found
seems to be more in accordance with the loose population and relatively lower extinction (A$_v$=7.0) found
in FSR249. This choice would put this object at a distance of 1.4-1.7 kpc which is somewhat closer than 
reported by K13 but compatible with distance estimates for the SFR DR21 by G12 and R12, reinforcing the 
idea that these two objects might be related. 

At this proposed distance, FSR249 is a parsec sized cluster ($\sim$ 2 pc diameter) with a 
well populated pre-main-sequence that appears to contain a high fraction of binaries. Its stellar population 
sums up to 310-460 M$_\odot$, with a distribution that barely follows the IMF, showing a consistent excess
of high mass stars down to $\sim$1 M$_\odot$. On the other hand, assuming the younger age range would 
put this cluster at 2.8-3.2 kpc with a total mass of 936-1375 M$_\odot$ and a diameter of 3.4 pc. The main 
charts involved in this cluster analysis are compiled in Fig.~\ref{fig:ap_fsr249}.

\subsection{LK13}
\label{sect.lk13}

LK13 resides at the southern tip of the "diamond ring", a gaseous ring-like structure powered by a central HII 
region, which appears to be connected to the DR21 (approximately 15\arcmin northeast) by filaments 
\citep{m04}. However, inspection of the composite Spitzer image shown in Fig.~\ref{fig:lk13} reveals that 
while LK13 appears to be illuminating and ionising the neighbouring southern and western gas structures, 
there is no such indication for the northern gas structure, which forms the diamond ring. 

This object was first identified by DB01 and later confirmed as a cluster candidate by LK02 which
assigned a minimum visual extinction of $\sim$9 and predicted the presence of 12 B stars in the cluster 
region, assuming a distance compatible of the Cygnus OB2 association ($\sim$1.6 kpc). Subsequent work 
by \citet{b10} has acquired spectra of 536 massive star candidates in Cygnus-X, finding only two nearby 
B-stars, largely outside the projected radius of LK13, but also reporting two very bright K0 stars inside its 
tidal radius, one being in the very centre of LK13.

There is little doubt that LK13 is a real, emerging (A$_v$$\sim$7) cluster. Although our analysis have found 
two equally probable age ranges of 4-6 Myr and 9-13 Myr, the prevalent gas content points to the younger 
value which corresponds to a distance of 450-550 pc, much closer than previously thought. Based on these 
results we estimate that the most massive stellar object near the centre of this cluster has 2-4 M$_\odot$, 
which is in rough agreement with its early K-type spectra found by \cite{b10}, if one considers the high binary 
fraction expected to be found in the central region of the cluster. 

These results would picture it as a compact cluster ($\sim$ 0.5 pc diameter) in a foreground layer of the 
complex associated with the Cygnus Rift molecular cloud, found at a distance of 500-800 pc by G12. Its 
stellar population sums up to 40-45 M$_\odot$ and appears to follow the IMF, albeit unreliable at 
the high mass end due to poor statistics. 
On the other hand, adoption of the older age range would lead to 
a much shorter distance of 320-400 pc, implying a smaller diameter (0.3 pc) and total mass (32-37 M$_\odot$).
The main charts involved in this cluster analysis are compiled in Fig.~\ref{fig:ap_lk13}.

\subsection{BDB60}

\begin{figure}
\centering
\includegraphics[width=\linewidth]{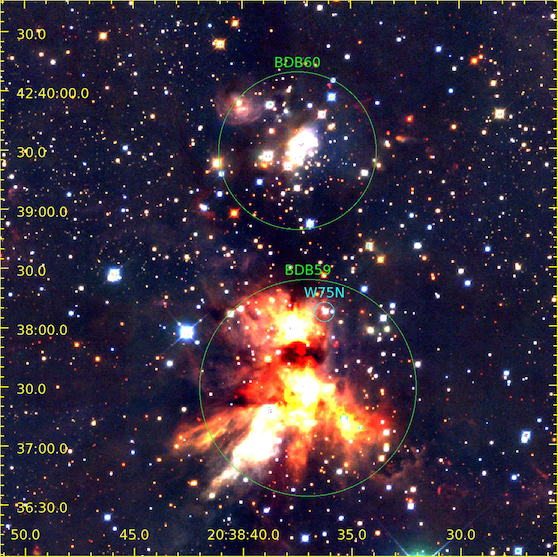} \\ \vspace{3mm}
\includegraphics[width=\linewidth]{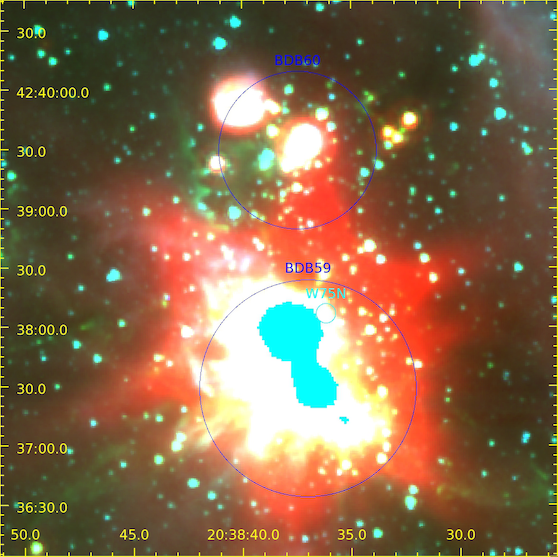}
\caption{Colour-composite images of the region around W75N, as seen from our $J$, $H$ and $K_s$ 
images (top) and from Spitzer $[3.6]$, $[4.5]$ and $[24] \umu m$ channels (bottom). The extraction radius 
for BDB59 and BDB60 are indicated in the figure.}
\label{fig:bdb60}
\end{figure}

This object was first identified as a cluster by DB01 and classified as a resolved pair to BDB59. 
However, even though it is separated from BDB59 by only $\sim$2\arcmin, it appears substantially different 
from its neighbour. Spitzer images shown in Fig.~\ref{fig:bdb60} reveals that it presents much less emission 
in the longer wavebands and that its stellar content is almost completely exposed, suggesting a more 
evolved evolutionary stage. Moreover, although BDB60 bright stars are projected very close to the very 
embedded BDB59, they show no evidence of interaction with its massive gas content, indicating they could 
reside in distinct layers (i.e. the foreground) of the cloud complex instead of being associated with BDB59.
Indeed, early CO mapping of this region by \citet{c77} has found gaseous structures associated with W75N 
that appears to belong to Great Cygnus Rift rather than being related to the DR21 complex to the south.

Due to the scarcity of stars and extinction heterogeneity, we could not reliably determine the physical 
parameters for this candidate cluster. Although we have found that its spatial distribution is consistent with a
king-profile with a tidal radius of 74\arcsec, close to the initial size estimate by DB01, the poor statistics 
precludes a definite conclusion on its nature, as such profile could very well arise from a random gathering of 
a few field stars.

\subsection{BDB59}

This stellar cluster is coincident with the well studied SFR W75N which has been 
resolved into at least three radio sources \citep{htft94}, possibly associated with ultra-compact HII 
regions and water masers found in the vicinity \citep{t97}. Recent observations in the mid infrared 
\citep{pts06} and millimetre \citep{skt04} wavebands have shown that these are probably excited by still 
embedded early B-type protostars aged from 1 to 5 Myr. Parallaxes measurements by R12 have constrained 
a distance of $1.30 \pm 0.07$ kpc for W75N while CO measurements by G12 has found an emission layer 
at 1.0-1.8 kpc for this SFR, possibly overlapping (but not interacting) with a background layer related to
DR21 in the south.

In order to reduce contamination from nearby BDB60 object (see Fig.~\ref{fig:bdb60}), we have limited our 
analysis to the inner 55\arcsec of this object. Although this aperture is large enough to encompass all radio 
and HII sources previously detected in this SFR, our photometry cannot reach the deepest, embedded sources,
thus being more biased towards the revealed, outer population. Nevertheless, we have estimated from its fitted
King-profile that this sample correspond to $\sim$88\% of the one at its tidal radius and have adjusted the 
derived total mass accordingly. 

Although still plagued by contamination and highly variable extinction, our analysis shows that most stars 
comprises a young (1-3 Myr), embedded (minimum A$_v$= 13.5) star cluster at a distance of 1.1-1.4 kpc. 
Using its tidal radius as a size indicator would result in an actual diameter of 1.4 pc. The total stellar mass 
amounts to 220-270 M$_\odot$ with a distribution does not seems to follow the IMF in the high-mass end, 
showing a excess of massive (m $>$ 5 M$_\odot$) stars. In this sense, the reduced radius used in the 
analysis could be effectively filtering out lower mass stars in the cluster outer region, if its stellar content is 
mass segregated. The main charts involved in this cluster analysis are compiled in Fig.~\ref{fig:ap_bdb59}.

\subsection{BDB58}

This massive emerging cluster was first identified by DB01 and lies close to the DR21(OH) source, located at 
the northern tip of a gaseous ridge that extends towards the main body of DR21, about 3\arcmin to the 
south. SCUBA dust-continuum mapping by \citet{d07} has found this gaseous structure extends further 
north into a filamentary structure, and suggested a likely distance of 3.0 kpc for the entire complex. 
Methanol maser kinematics by \citet{pmb05} constrains a similar distance for the northern ridge of DR21 
based on two sources, one of which lies inside BDB58 radius.

\begin{figure}
\centering
\includegraphics[width=\linewidth]{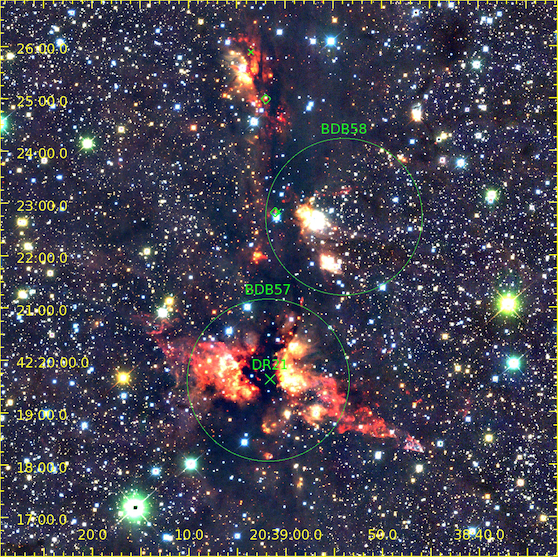} \\ \vspace{3mm}
\includegraphics[width=\linewidth]{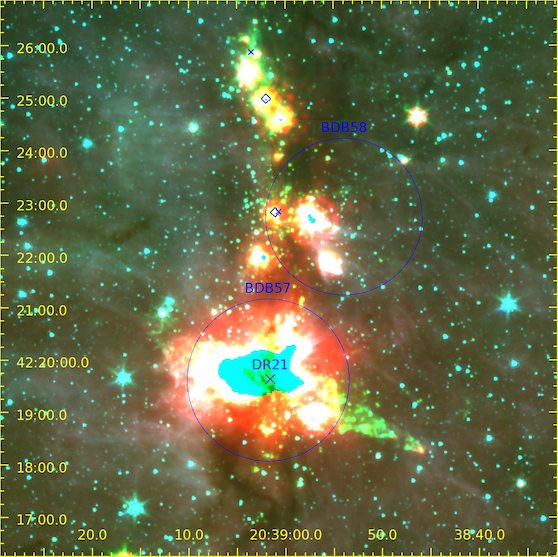}
\caption{Colour-composite images of the region around DR21, as seen from our $J$, $H$ and 
$K_s$ images (top) and from Spitzer $[3.6]$, $[4.5]$ and $[24] \umu m$ channels (bottom). The extraction 
radius for BDB57 and BDB58 are indicated, along with nearby masers used for distance estimation via 
parallax (diamonds) and kinematics (crosses).}
\label{fig:bdb57}
\end{figure}

Fig.~\ref{fig:bdb57} shows that although it appears more extended and completely embedded in the Spitzer 
image, part of its western side has already emerged revealing a very red stellar population in our 
near-infrared image. The ridge and gaseous filaments can be seen as dark lanes in our images.

Our results points out that this is an emerging cluster probably younger than 2.5 Myr, presenting a 
minimum A$_v$=13.5 and residing at a distance of 2.7-3.3 kpc. At this distance its tidal radius would 
translate into a diameter of 2.6 pc. Although an exponent is given, its stellar mass distribution does not 
appear to follow the IMF at high mass end (m $>$ 1 M$_\odot$) showing an excess of high mass stars, 
and summing up to 610-770 M$_\odot$. The main charts involved in this cluster analysis are compiled in 
Fig.~\ref{fig:ap_bdb58}.

\subsection{BDB57}

This nascent cluster, often identified as W75S or DR21 IR cluster \citep{h94}, lies in the centre of DR21, the 
most studied SFR in the Cygnus-X north complex. However, although a multitude of radio and IR studies 
revealed several clumps, molecular jets, masers and YSO's in this region, a definitive distance is still lacking.
Many early studies \citep[e.g.][]{c82} and more recent ones \citep[e.g.][]{d07} have been favouring distances 
up to 3 kpc, largely based on kinematical data and galaxy rotation models. Recent maser parallaxes by R12 
have found 1.5 kpc, based on three sources residing largely in the northern ridge (see Fig.~\ref{fig:bdb57}), 
while CO emission maps by G12 puts it between 1.5-2.5 kpc. 

Presenting the largest extinction found among our sample (minimum A$_v$=24.4), we have only been able
to provide tentative parameters for BDB57, given its scant population and large extinction heterogeneity. 
The most probable age found ($\sim$1.4 Myr) is very close to the minimum age sampled by our method 
(1.0 Myr), suggesting that it could have been limited by our sampling. The corresponding distance 
(assuming a 1.0-1.4 Myr age) is 600-800 pc and should be regarded as a minimum threshold since the 
adoption of a younger age would certainly reflect in an increased distance modulus (see Fig.~\ref{fig:ext}).
Based on these parameters, the cluster diameter would be 0.8 pc with a stellar population amounting to 
160-220 M$_\odot$ that appear to follow the IMF in the low mass range, but not in the high mass 
(m $>$ 1 M$_\odot$) regime. Again, given the distance uncertainty, these should be regarded as minimum 
thresholds.

Although our tentative distance modulus is much smaller than previous findings, it's not unrealistic to expect 
an increase of one magnitude (leading to a distance of $\sim$1.3 kpc) for a 0.6 Myr aged system. 
However, given our data, we find it impossible to reach values as high as 2.0 kpc. In addition, previous distance 
estimates for the BDB57 cluster are not very well constrained as they were either carried over globally, 
through the gas/dust component of the whole DR21 complex, or punctually, over a few maser sources at least 
4\arcmin away from the stellar cluster centre, well outside its derived tidal radius. In fact, \citet{hlr02} has 
found that two radio sources in the very center of DR21, for which no distance estimate has ever been made, 
have no counterpart in the NIR, presenting only extended nebular emission that does not originate in the radio 
sources either. They conclude that these sources might be completely unrelated to the star formation traced in 
the NIR. Since we cannot say if this is the case for the BDB57 cluster, the derived cluster parameters of this 
cluster remains tentative ones, at best.
The main charts involved in this cluster analysis are compiled in Fig.~\ref{fig:ap_bdb57}. 

\subsection{Bica3}

This object was first classified as a possible distant and old physical system by \citet{bbd03} based on the 
identification of a CMD sequence that resembles a giant branch. Although featureless on Spitzer images, 
our images show Bica3 as a rich and compact population of faint stars approximately 15\arcmin east of 
DR21 complex. 

Our analysis indicates that Bica3 is indeed a very reddened (A$_v$=15.1) and old (450-630 Myr) open 
cluster probably residing at the background Perseus arm (distance 4.5-5.0 kpc), corroborating the initial 
guess on this object nature. These parameters would imply a diameter of 7.7 pc at its tidal radius and a 
total stellar mass of $\sim$3000 M$_\odot$, leading to a surprisingly big open cluster for its age. Its mass 
distribution shows a very flat spectrum, probably due to leftover contaminants at brighter magnitudes and/or
incompleteness at the fainter end.

We should note that the total mass for Bica3 is probably overestimated. The actual observed 
mass function sums up to $sim$1600 M$_\odot$ sampling the mass range between 1-3 M$_\odot$. The 
extrapolation of the mass function to lower masses to account for the 'unobserved' stars was done 
using the fitted power law slope of 1.3$\pm$0.3 rather that the canonical value of 2.3, and can be highly 
dependent on the normalisation adopted and the involved star counts. A leftover field population would 
be substantially enlarged when extrapolating star numbers into the low mass domain. In that matter, although 
the field decontamination procedure employed could successfully account for the abundant population in the 
local arm, its performance in sampling and removing the obscured field population of the Perseus arm in the 
background is much more reduced. The main charts involved in this cluster analysis are compiled in 
Fig.~\ref{fig:ap_bica3}.

\subsection{LK15}

\begin{figure}
\centering
\includegraphics[width=\linewidth]{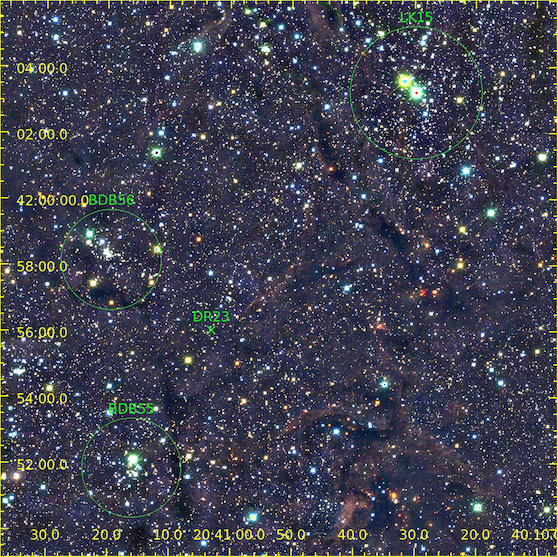} \\ \vspace{3mm}
\includegraphics[width=\linewidth]{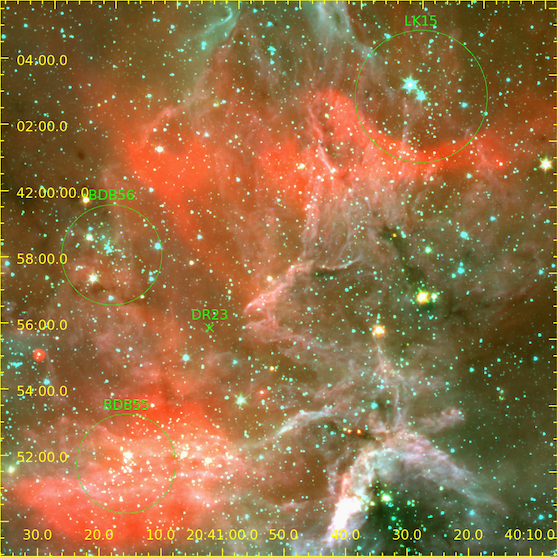}
\caption{Colour-composite images of the region around DR23 (marked with an X), as seen from our $J$, 
$H$ and $K_s$ images (top) and from Spitzer $[3.6]$, $[4.5]$ and $[24] \umu m$ channels (bottom). 
The extraction radius for LK15, BDB55 and BDB56 are indicated.}
\label{fig:lk15}
\end{figure}

This object was identified by LK02 and reported as being possible associated with DR23, 
approximately 10\arcmin southeast. Although a stellar density enhancement is discernible in our NIR image 
(Fig.~\ref{fig:lk15}), specially at brighter magnitudes, it is hard to assign a clear circular border to this object 
as the presence of various foreground dust lanes gives it a very structured appearance in the NIR. Since 
King-profiles could not reliably describe its stellar density profile, we have instead turned to the brighter stellar 
content ($K_s$ $<$ 15.5) density charts to derive its structural parameters. Although the resulting parameters 
are likely to represent the higher mass stellar content, our derived tidal radius ($\sim$ 2\arcmin) is considerably 
smaller than the $\sim$7\arcmin adopted by LK02, which is large enough to include several gaseous filaments 
and nearby dark ridges along with the exposed stellar group. 

Despite its youth (1.0-3.0 Myr) LK15 appears to be an emerged population (A$_v$=11.5) at a distance 
between 1.7-2.1 kpc, consistent with the estimated value for DR23 \citep[2.0 kpc,][]{os93}. One very bright 
star at its centre and presenting compatible extinction (there is a second bright, but much redder star) could 
be responsible for clearing away the gas and producing an ionisation front at its south border, seen in the 
Spitzer images. Our derived parameters and the colour calibrations by \citet{k00} suggests it to be a late 
O-type star and the used isochrone tables gives a $\sim$70 M$_\odot$ star. In addition, our results imply a 
physical diameter of 2.0 pc and an expected stellar mass of 370-485 M$_\odot$, showing a small excess 
of high mass stars (m > 1 M$_\odot$) when compared to the IMF.  The main charts involved in this cluster 
analysis are compiled in Fig.~\ref{fig:ap_lk15}.

\subsection{BDB55}

This candidate cluster was identified by DB01 and assigned as being probably associated 
with the DR23 radio/HII region approximately 4\arcmin northwards. Since the work by LK02, it has been 
misidentified in the SIMBAD database as their object 15 (LK15 in this work), located $\sim$15\arcmin 
to the northwest. It has also been processed by K13 (object \#3395), deriving an age of 2.5 Myr, A$_v$ 
$\sim$ 6.5 and distance of 4.2 kpc. However, we find it hard to describe any stellar population in Cygnus with 
these parameters since a much higher reddening would be expected from such young, distant stellar group. 

Spitzer imaging of BDB55 (Fig.~\ref{fig:lk15}) shows that it is characterised by a small clumping of bright 
stars that appears to be right inside a filamentary, ionised structure extending from a gaseous complex in 
the west. However, our NIR images reveals that the bright stars that compose this object are in fact, 
foreground blue sources, superimposed on a very reddened population inside the gaseous structure. 

In fact, we have found that this object most likely does not correspond to a physical system. Its does
not show any density enhancement over the background in the fainter magnitude domain and its bright star
clumping does not correspond to a King-profile distribution. Moreover, the faint stellar population in the area
are very similar to that of the nearby field, being completely decimated in the our field removal analysis.
The leftover bright stars showed very small extinction and a vertical structure on the CMD typical of the 
upper main-sequence stars. 

\subsection{BDB56}

This object was found by DB01 and was also assigned to be possibly related to the nearby DR23 
region, about 3\arcmin southwest. Processing by K13 (object \#3396) have derived an age of 800 Myr, 
A$_v$ $\sim$ 5 and distance of 1.2 kpc. For such small stellar numbers, these parameters would 
correspond to a remnant or dissolving old cluster among Cygnus predominantly young population. Although 
possible, we believe their results might have been misguided by the very reddened (but still young) field 
stars. 

Fig.~\ref{fig:lk15} shows BDB56 as a very noticeable star clustering in both the Spitzer and our NIR 
images. Dark lanes are also visible in the cluster vicinity, probably associated with foreground clouds in 
Great Cygnus Rift as they do not seem to be interacting in any way with the local, bright stellar content. 
Overall, the intra-cluster region appears to be devoid of gas, suggesting a more evolved evolutionary 
state. A bright blue star near its centre has been spectral typed as a A7 star with very small extinction 
(A$_v$=3.9) by \citet{b10}. Our photometry and the colour calibrations by \cite{k00} gives this star a 
distance of approximately 630 pc, compatible that of the foreground Cygnus Rift. 

Our analysis have shown that BDB56 is a young (age 2-5 Myr), emerged cluster (A$_v$=9.8) at a 
distance of 1.4-1.6 kpc, indicating that it might not be associated with DR23 if a 2.0 kpc distance is 
assumed for this radio source. At our derived distance, this cluster has 1.2 pc diameter with a total stellar 
mass of 140-190 M$_\odot$ with a distribution compatible with the IMF, albeit showing an slightly excess 
of massive stars. Furthermore, our results points out that brightest star in the vicinity has reddening 
compatible with the cluster population and is probably a B0-B1 star according to the color calibrations by 
\citet{k00}. The main charts involved in this cluster analysis are compiled in Fig.~\ref{fig:ap_bdb56}.

\section{General results and discussion}

\subsection{Embedded cluster properties}

\begin{figure}
\centering
\includegraphics[width=\linewidth]{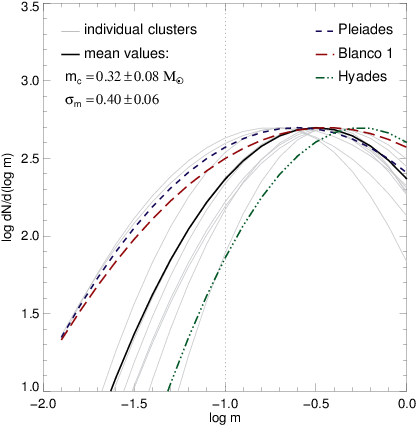}
\caption{Comparison of the individual (thin lines) and mean (thick line) log-normal mass functions derived in 
this work with the reference values (dashed lines) for the Pleiades, Blanco1 and Hyades clusters. The vertical
dotted line represent the minimum mass sampled in our IMFs (0.1 M$_\odot$).}
\label{fig:mf}
\end{figure}

Concerning the low-mass, log-normal domain of the IMF (m $<$ 1 M$_\odot$), the clusters in this region 
presented similar mass distributions, showing a mean characteristic mass of $m_c$=0.32$\pm$0.08 
M$_\odot$ and a mean mass dispersion of $\sigma_m$=0.40$\pm$0.06. Fig.~\ref{fig:mf} compares their 
individual mass functions with reference values from the Pleiades, Hyades and Blanco 1 clusters 
\citep{m03,m07}. It is clear that the IMF characteristic mass appears to be robustly constrained around 
0.3 M$_\odot$, while the mass dispersion appears to present larger variations between the different regions. 
In this sense, our mass distributions are 
narrower than those found for the Pleiades and Blanco 1, ($\sigma_m$=0.52-0.58) but agrees very well with 
the value derived for the old Hyades cluster ($\sigma_m$=0.38). However, when compared to these clusters, 
our sampled masses are generally one decade shallower presenting only one or two reliable mass bins fainter
than the log-normal peak (i.e. the characteristic mass). A proper consideration of the brown dwarf population 
in each cluster should be done to better constraint these parameters, particularly the mass dispersion. 

By analysing the clusters structural parameters, we have found that the sizes of all young objects are well 
constrained at the one parsec scale, with most diameters ranging from $\sim$1 pc to 2 pc. Although
being considerably smaller than evolved open clusters, their derived 
concentration parameters (2.0 $\le$ $r_t$/$r_c$ $\le$ 4.0) agree very well with the values typically observed 
in these older clusters \citep[e.g.][]{sa09}. In addition, taking into account their masses and assuming a 
virialised population, their calculated velocity dispersions (i.e. $\sigma_v$=$\sqrt{GM/r}$) amount to 
$\approx$2 km/s for all clusters, also in agreement with values generally found in older open clusters. 
This is remarkable, considering that the we are sampling much younger clusters distant from 0.5 to 3.0 
kpc across at least three different SFRs, and puts strong constraints into their very early structural 
configuration. 
 
In addition, the derived mean stellar densities ($\bar{\rho}$) seems to indicate that the older star clusters 
(age $>$ 4 Myr) are, on general, slightly denser than the younger ones of the same mass range (e.g. 
LK12, FSR249 and LK15), contrary to the expected evolutionary behaviour of stellar populations. However, 
given the very early age of our youngest clusters, it could be wondered whether they are fully formed 
systems or if one could still expect an enlargement of their stellar population due to ongoing star formation,
accretion from the nearby SFR or simply emerging from its parental cloud as the gas is dissipated. In this 
sense, it is clear from our derived extinctions that most of our clusters under $\sim$ 4 Myr are still embedded 
and thus able to enhance their overall stellar densities over the next few Myr.

An observed trend between the clusters age and mass function slope at higher masses (m $>$ 1 M$_\odot$)
might be related to a similar issue. The younger clusters present a shallower slope with respect to the IMF while 
older ones show steeper distributions, compatible with the IMF nominal value ($\alpha$=2.35).
Given the early age and highly embedded state of the younger clusters, it is unlikely that they have 
lost their lower mass content due to dynamical evolution. 
Therefore, it could be argued that these clusters populations either do not follow the IMF or were not 
sampled at a appropriate spatial scale to properly populate the observed IMF.
In this sense, it is possible that some of their low mass content, formed in its outer regions, 
might not have had the time to sink to their central regions where our stellar samples were extracted. Indeed, 
consideration of the large gas content inside these objects would greatly increase their (stellar derived) tidal 
domain, allowing them to draw stars from much farther away on the SFR into the forming cluster. 
This would support a scenario where the massive stars and a population of low
mass stars are formed in densest, central region of the cluster, while some additional lower mass content 
would be formed in the outer filamentary structure and then accreted to the cluster centre over its first few 
Myr, up to the gas expulsion phase.
At this point, loss of the gaseous content would greatly reduce the cluster 
gravitational potential halting the stars accretion and leading to a rapid expansion. This appears to be confirmed 
by our older cluster group, with ages greater than 4 Myr, presenting slightly larger stellar densities and steeper 
MF slopes that are more compatible with the IMF.

A similar scenario have been recently found concerning the star formation in the DR21 ridge (our BDB57 
cluster) and its nearby filamentary structure by \citet{h12} based on the local gas dynamics and {\it Herschel} 
maps. The gravitationally unstable outward filaments forms cores and protostars which flows, along with the 
gas, towards the the central ridge (the cluster), where most of the densest and more massive cores resides. 
This behaviour was also observed in the gas content of other star forming regions \citep{s13}, but has yet to 
be observed on the stellar content of a nascent cluster.

\subsection{Cygnus-X North}

By examining our extinction distributions across our clusters, we have found that the extinction towards the 
Cygnus-X north presents a stable minimum of about 5-6 mag. towards the more evolved clusters in 
the complex, possibly amounting to the foreground gaseous content. 
When accounting for this foreground layer, the local extinction in this complex appears to amount to just 1-2 
mag. for the evolved clusters (age $>$ 4 Myr) while spanning values between 5-7 mag. in most of the 
younger ones. Interestingly, even though the older clusters presented slightly sharper extinction distributions, 
this difference was found to be small across our clusters, as they showed about the same extinction 
dispersion (evaluated as the FWHM of the distribution) of about 2-3 mag. around their peak value. Since this 
dispersion generally represents the intra-cluster variation of the extinction column, its somewhat constant 
value might be tied to similar spatial depths shared by these clusters.

Estimated distances to these clusters appears to confirm previous segregation of the gaseous structure
of this region into three layers \citep{g12} at distances of $\sim$500-950 pc (LK12, LK14 LK13, 
BDB57), $\sim$1.4-1.7 kpc (FSR249, BDB59, LK15, BDB56) and $>$2.5 kpc (BDB58). 

The estimated distances to the clusters belonging to the second layer, particularly that of BDB56, 
the southernmost cluster in our sample (1.4-1.6 kpc), agree very well with recent estimates to Cygnus-OB2 
association \citep[e.g 1.4 kpc -][]{h03}, pointing out to a possible common origin between the association and 
southern part of Cygnus-X complex. Given the youth of this cluster (1-4 Myr) and the relatively small distance 
to the association ($\sim$40 pc at 1.4 kpc), a triggered formation scenario is not unlikely considering that 
Cygnus OB2 could be aged up to 10 Myr. 

\section{Conclusion}
We have performed an unprecedentedly deep survey of the Cygnus-X north complex in the near-infrared 
unveiling a rich population of low mass stars and embedded stellar clusters in this region. By developing a tool 
to properly account for the complex extinction pattern in the region and infer the most probable age and 
distance of a stellar population, we have been able to uniformly characterise a dozen mostly unstudied clusters 
throughout this region. 

Our results points out that these populations were likely born inside dense, 
embedded stellar structures of about one parsec size and presenting slightly shallower mass functions slopes 
than the IMF, thus presumably lacking some of its lower mass content. Although by themselves the slopes 
provides evidence against the universality of the IMF, additional cluster parameters (i.e. age, stellar density) 
appear to support a scenario were these very young clusters are likely to accrete additional low mass stars 
formed in their vicinity during their first few Myr, reaching the age of 4 Myr with a fully fledged IMF inside their tidal 
radius. The gas expulsion phase at about 3 Myr would play an important role in halting this process. 

Fittings of log-normal mass functions at the low mass domain 
(m < 1 M$_\odot$) points out that the general system IMF in this region appears to have a fairly robust 
characteristic mass (m$_c$=0.32$\pm$0.08), that is compatible with those found in other regions, but presents 
a narrower mass dispersion ($\sigma_m$ = 0.40 $\pm$ 0.06) with respect to other young clusters.

Clustered star formation appears to be happening in Cyngus-X north in three different layers, located at 
distances of 500-950 pc, 1.4-1.7 kpc and 3.0 kpc, consistent with previous distances estimates for 
The Great Cygnus Rift, the W75N region and the DR21 filament, respectively. 
Finally, at least one young cluster at the southern 
part of Cygnus-X north has distance compatible with that of the Cygnus-OB2 association, implying that some 
interaction may have taken place between these two regions.

\section*{Acknowledgments}
The authors would like to thank the anonymous referee for the suggestions and critics that contributed to
greatly enhance this work.
The authors acknowledge the support of the brazilian funding agencies CAPES, CNPq. This work was funded in 
part by the French national research agency through ANR 2010 JCJC 0501-1 ``DESC'' (PI E. Moraux).
Based on observations obtained with WIRCam, a joint project of CFHT, Taiwan, Korea, Canada, France, 
and the Canada-France-Hawaii Telescope (CFHT) which is operated by the National Research Council 
(NRC) of Canada, the Institute National des Sciences de l'Univers of the Centre National de la Recherche 
Scientifique of France, and the University of Hawaii. 
This research has made use of the NASA/ IPAC Infrared Science Archive, which is operated by the Jet 
Propulsion Laboratory, California Institute of Technology, under contract with the National Aeronautics and 
Space Administration. 
This publication makes use of data products from the Two Micron All Sky Survey, which is a joint project of 
the University of Massachusetts and the Infrared Processing and Analysis Center/California Institute of 
Technology, funded by the National Aeronautics and Space Administration and the National Science Foundation.
This research has made use of the SIMBAD database and of the VizieR catalogue access tool, operated at 
CDS, Strasbourg, France








\appendix

\section{Analysis charts}
\begin{figure*} 
\centering
\begin{sideways}
\begin{minipage}{230mm}
\includegraphics[width=16cm,angle=90]{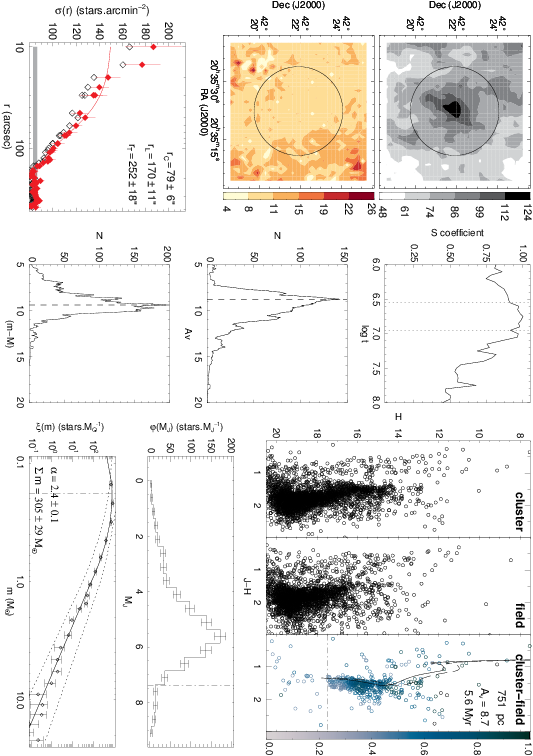}
\caption{LK12 analysis charts. Top-left: general density map showing the cluster extraction radius (solid line).
Centre-left: extinction map of the cluster vicinity. Bottom-left: RDP showing the derived structural parameters.
Top-middle: $S$ coefficient (solid line) showing the adopted age interval (vertical dotted lines). 
Bottom-middle: extinction and distance modulus distribution from which their expected values (vertical 
dashed lines) were derived. Top-right: CMDs comparing the cluster, field and decontaminated samples; 
single-star (solid line) and same mass binary (dashed line) isochrones with the most probable physical 
parameters are shown. Bottom-right: derived LF showing the MF fitting limit (vertical dot-dahsed line) 
and derived mass distribution showing the best MF fit (solid line) and the expected IMF domain (dotted lines) .}
\label{fig:ap_lk12} 
\end{minipage}
\end{sideways}
\end{figure*}

\begin{figure*} 
\centering
\begin{sideways}
\begin{minipage}{230mm}
\includegraphics[width=16cm,angle=90]{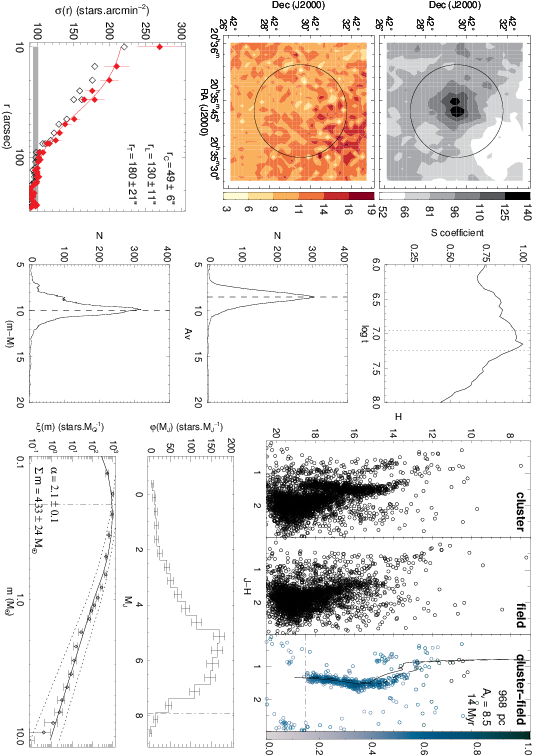}
\caption{LK14 analysis charts. Panels are the same as in Fig.~\ref{fig:ap_lk12}.}
\label{fig:ap_lk14} 
\end{minipage}
\end{sideways}
\end{figure*}

\begin{figure*} 
\centering
\begin{sideways}
\begin{minipage}{230mm}
\includegraphics[width=16cm,angle=90]{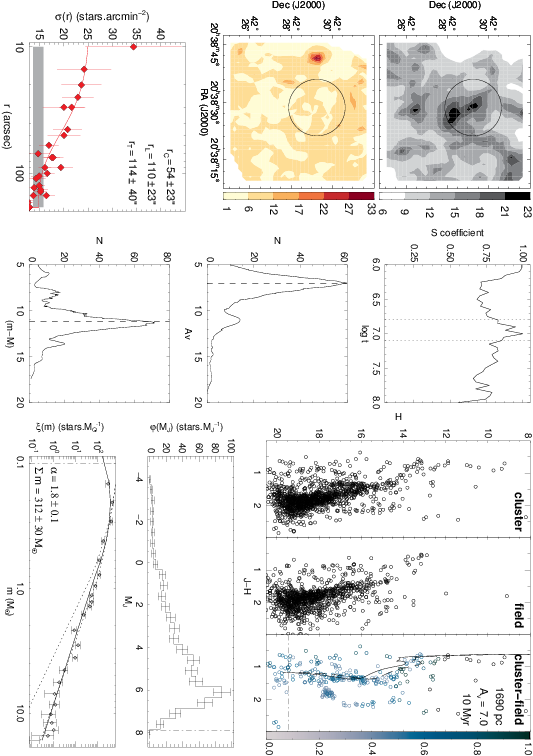}
\caption{FSR249 analysis charts. Panels are the same as in Fig.~\ref{fig:ap_lk12}.}
\label{fig:ap_fsr249} 
\end{minipage}
\end{sideways}
\end{figure*}

\begin{figure*} 
\centering
\begin{sideways}
\begin{minipage}{230mm}
\includegraphics[width=16cm,angle=90]{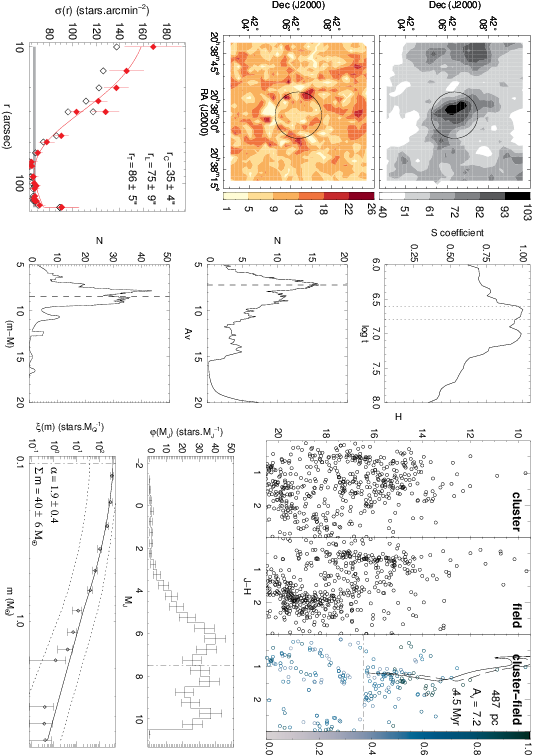}
\caption{LK13 analysis charts. Panels are the same as in Fig.~\ref{fig:ap_lk12}.}
\label{fig:ap_lk13} 
\end{minipage}
\end{sideways}
\end{figure*}

\begin{figure*} 
\centering
\begin{sideways}
\begin{minipage}{230mm}
\includegraphics[width=16cm,angle=90]{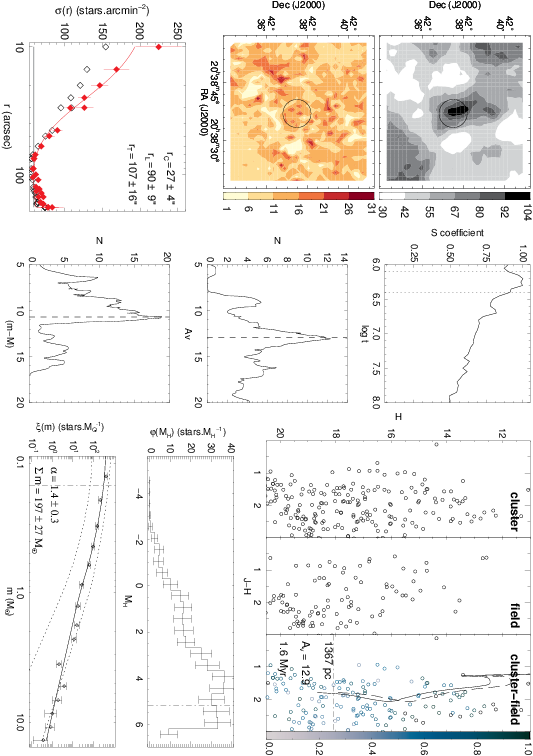}
\caption{BDB59 analysis charts. Panels are the same as in Fig.~\ref{fig:ap_lk12}.}
\label{fig:ap_bdb59} 
\end{minipage}
\end{sideways}
\end{figure*}

\begin{figure*} 
\centering
\begin{sideways}
\begin{minipage}{230mm}
\includegraphics[width=16cm,angle=90]{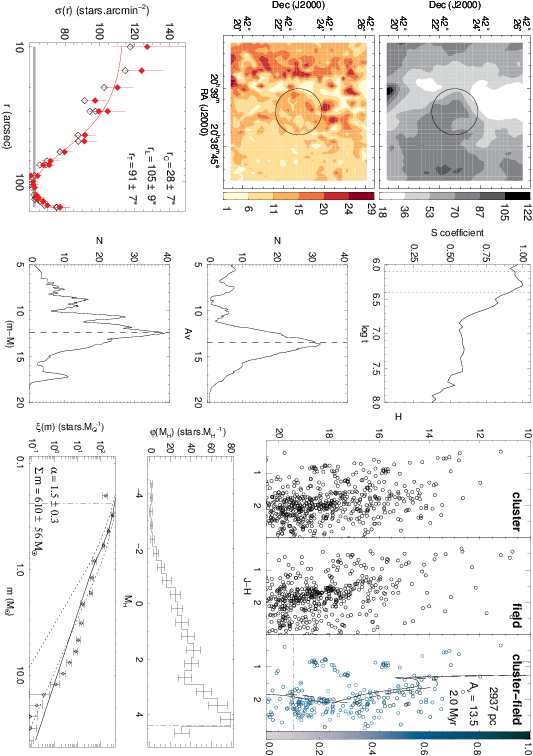}
\caption{BDB58 analysis charts. Panels are the same as in Fig.~\ref{fig:ap_lk12}.}
\label{fig:ap_bdb58} 
\end{minipage}
\end{sideways}
\end{figure*}

\begin{figure*} 
\centering
\begin{sideways}
\begin{minipage}{230mm}
\includegraphics[width=16cm,angle=90]{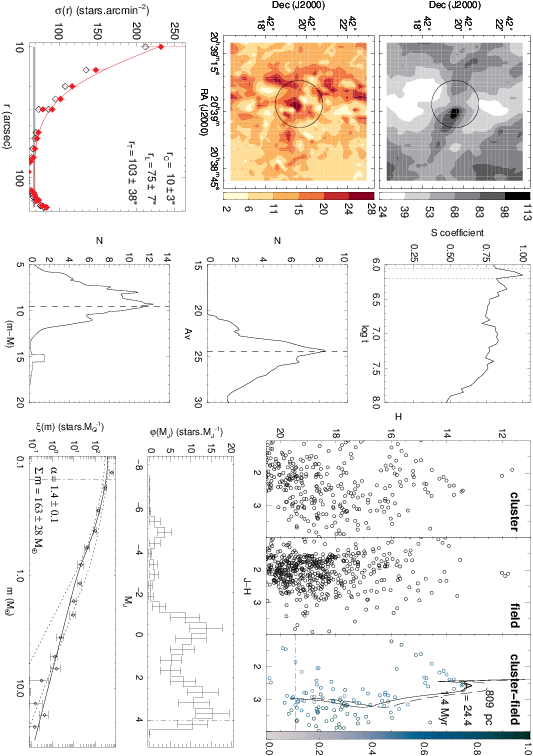}
\caption{BDB57 analysis charts. Panels are the same as in Fig.~\ref{fig:ap_lk12}.}
\label{fig:ap_bdb57} 
\end{minipage}
\end{sideways}
\end{figure*}

\begin{figure*} 
\centering
\begin{sideways}
\begin{minipage}{230mm}
\includegraphics[width=16cm,angle=90]{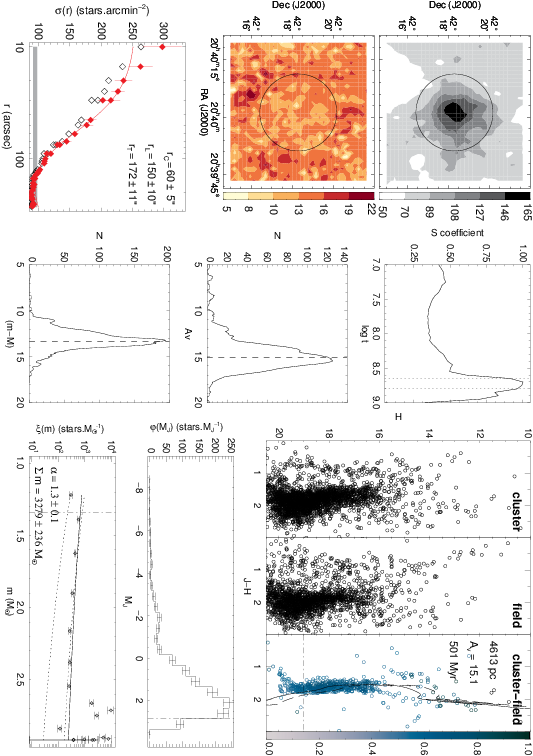}
\caption{Bica3 analysis charts. Panels are the same as in Fig.~\ref{fig:ap_lk12}.}
\label{fig:ap_bica3} 
\end{minipage}
\end{sideways}
\end{figure*}

\begin{figure*} 
\centering
\begin{sideways}
\begin{minipage}{230mm}
\includegraphics[width=16cm,angle=90]{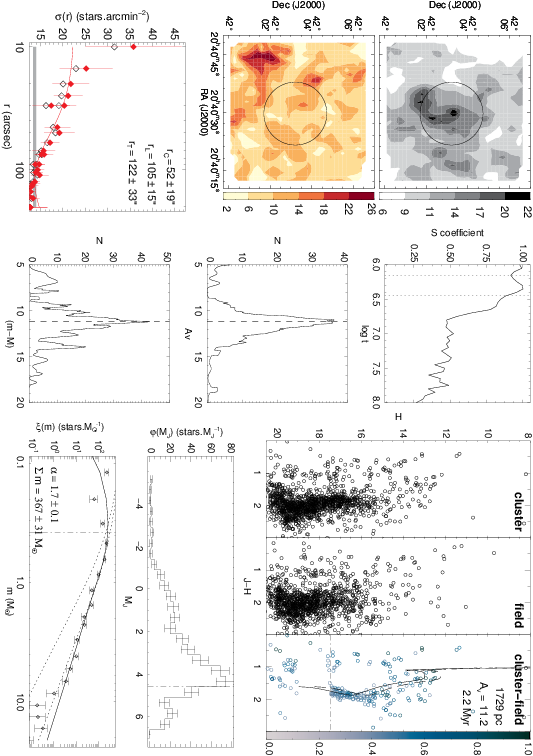}
\caption{LK15 analysis charts. Panels are the same as in Fig.~\ref{fig:ap_lk12}.}
\label{fig:ap_lk15} 
\end{minipage}
\end{sideways}
\end{figure*}

\begin{figure*} 
\centering
\begin{sideways}
\begin{minipage}{230mm}
\includegraphics[width=16cm,angle=90]{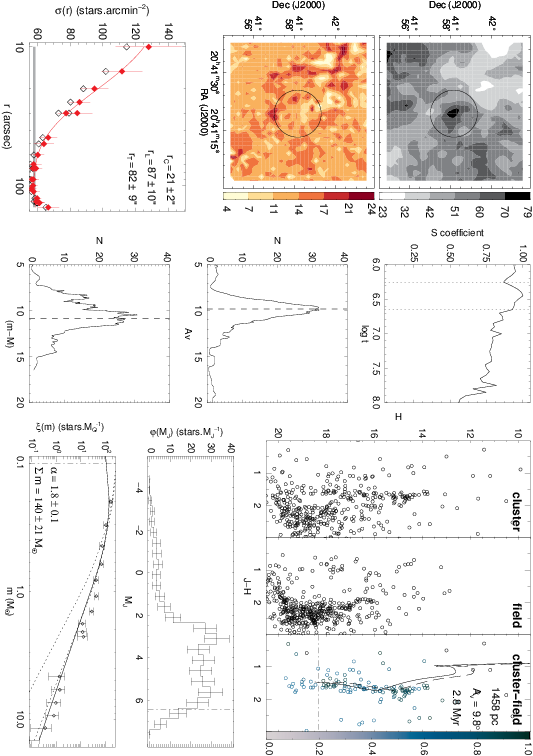}
\caption{BDB56 analysis charts. Panels are the same as in Fig.~\ref{fig:ap_lk12}.}
\label{fig:ap_bdb56} 
\end{minipage}
\end{sideways}
\end{figure*}

All the plots and figures generated in the analysis of our clusters were compiled into full page charts, presented
in Figs.~\ref{fig:ap_lk12}-\ref{fig:ap_bdb56}, through this appendix. 

\bsp	
\label{lastpage}
\end{document}